\documentclass{mn2e}
\input{epsf}
\newcommand{\be}{\begin{eqnarray}}
\newcommand{\ee}{\end{eqnarray}}

\begin{document}

\title{Beyond the $\Lambda$CDM cosmology: complex
composition of dark matter.}
\author[Demia\'nski \&  Doroshkevich ]
       {M. Demia\'nski$^{1,2}$,  A.G. Doroshkevich$^{3}$,\\
        $1$Institute of Theoretical Physics,
                       University of Warsaw,
                       00-681 Warsaw, Poland\\
        $2$Department of Astronomy, Williams College,
           Williamstown, MA 01267, USA\\
        $3$Astro Space Center of Lebedev Physical
           Institute of  Russian Academy of Sciences,
                        117997 Moscow,  Russia\\
}

\date{Accepted ...,
      Received ...,
        in original form ... .}
\maketitle

\begin{abstract}
The mass and composition of dark matter (DM) particles
and the shape of the power spectrum of density
perturbations are estimated using recent observations
of the DM dominated relaxed objects -- dSph, THINGs
and LSB galaxies and clusters of galaxies. We consider
the most extensive available sample of observed objects with 
masses $10^6\leq M_{vir}/M_\odot\leq 10^{15}$ which
includes $\sim 60$ DM dominated galaxies and $\sim 40$
clusters of galaxies. We show that the observed
characteristics of these objects are inconsistent with
expectations of the standard $\Lambda$CDM cosmological
model. However, they are well reproduced by a mixed
CDM+WDM model with a significant contribution of the
HDM--like power spectrum with a relatively large damping
scale. We show that the central pressure of DM dominated
objects is surprisingly weakly dependent upon their
virial mass but it is very sensitive to the efficiency
of cooling of the baryonic component. In contrast, the
central entropy of both DM and baryonic components
strongly depends upon the virial mass of halo and the
period of halo formation. Unfortunately the available
data prevent our qualitative approach to reach more
reliable and definite conclusions which requires
confrontation of more representative observational data
with high resolution numerical simulations.
\end{abstract}


\section{Introduction}

The nature of dark matter (DM) particles is one of the
intriguing questions of modern physics. These particles are
an important element of the Standard Cosmology, they
represent $\sim 20-25\%$ of the mean matter-energy density
and explain some observed properties of the Universe (see,
e.g., Komatsu 2011; Larson 2011; Burenin \& Vikhlinin 2012;
Saro 2013; Ade et al. 2013; Samushia et al. 2014). At the
same time various candidates of DM particles are widely
discussed as a very important element of high energy physics.
This dual role of DM particles (see, e.g., Rubakov 2011)
explains the great attention which is recently devoted to
these problems.
Many possible candidates of the DM particles are now considered.
Thus, these particles may have masses ranging from massive
gravitons with $m\sim 10^{-19}eV$ and up to the supersymmetric
WIMPs with $m\sim 10^{13}GeV$. So wide range of possible
masses is a result of very weak observational restrictions
and implies similar wide range of other particle properties.
In particular it is possible to note specially such traditional
candidates as axino (Choi, Kim, Roszkowski 2013), or black
holes (Carr 2014), and such exotic ones as Atomic DM
(Cyr-Racine \& Sigurdson, 2013) and the flavor-mixed two
component DM models (Medvedev 2014) or reincarnation of
massive neutrino models (Costanzi et al. 2013;
Villaescusa-Navarro et al. 2013). Very detailed discussion of
various aspects of contribution of neutrinos to dark matter
in the context of latest observations of Planck mission, baryonic
oscillations and cluster properties can be found in Verde et
al. (2013) and Wyman et al. (2014).

In turn the more and more refined observations determine
evolution in time of the DM models. Historically in cosmology
DM particles were introduced in Doroshkevich et al. (1980)
and Bisnovaty-Kogan \& Novikov (1980), as the Hot Dark Matter
(HDM) model with the massive neutrino as the DM particles.
Soon after the DM models were extended by introduction
of the cold DM (CDM) and warm (WDM) models (Bond, Efstathiou
\,\& Silk, 1980; Bond \& Szalay, 1983; Primack, 1984;
Blumenthale et al., 1984; Bardeen et al. 1986) and even more
complex models of multicomponent (MDM) and unstable DM
particles (UDM) (Doroshkevich, Khlopov, 1984; Turner,
Steigmann, Krauss, 1984; Doroshkevich, Khlopov, Kotok, 1986;
Doroshkevich, Klypin, Khlopov, 1988).

All these models were solving some actual cosmological
problems but their potential was always limited and none
of them survived confrontation with observations. The scientific
progress generates more problems and poses new questions
what requires continuous modifications and development of
new models of DM particles. Thus, early in this century
observations of CMB fluctuations by the WMAP mission, SPT
and other ground telescopes established the $\Lambda$CDM
model as the best cosmological model (Bennet et al. 2003;
Komatsu 2011; Larson 2011; Saro 2013). This inference was
supported by Planck measurements (Ade et al. 2013),
observations of clusters of galaxies (see, e.g., Burenin \&
Vikhlinin 2012) and baryonic oscillations (Eisenstein \&
Hu 1998; Meiksin et al. 1999; Samushia et al., 2014). At
that time the main hope of identifying the ``missing''
cosmological components had been focused on links of possible
distortions of the kinetic of recombination and the
corresponding CMB fluctuations (see, e.g., Peebles et al.
2000; Doroshkevich et al. 2003).

On the other hand, for many years the emerging conflict between
the Standard $\Lambda$CDM theory and observations of clustering
on subgalactic scales is widely discussed. First, it is believed
that the $\Lambda$CDM model predicts an excess of low--mass
satellites of Milky Way, next is the core--cusp problem seen
as a discrepancy between the observed and simulated shape of
the density profiles in central regions of relaxed
objects (see, e.g., Bovill \& Ricotti, M., 2009; Koposov et
al., 2009; Walker, \& Penarrubia, 2011; Boylan-Kolchin et
al., 2012; Penarrubia et al. 2012; Governato et al. 2012;
Sawala, 2013; Teyssier et al. 2013; Laporte et al. 2013;
Collins et al. 2014). Significance of these conflicts is quite
moderate as objects with very different masses, densities and
evolutionary histories are compared (see, e.g., Penarrubia et
al. 2008). Moreover limited reliability of these contradictions
is enhanced by limited resolution of both the observations and
simulations (see, e.g., Mikheeva, Doroshkevich, Lukash 2007;
Doroshkevich, Lukash \& Mikheeva 2012; Pilipenko et al. 2012).

During last years the analysis of the $Ly-\alpha$ forest
becomes very popular again (see, e.g. Boyarsky at al. 2009a,b,
d; Dipak et al. 2012; Viel et al. 2013; Marcovi$\breve{c}$ \&
Viel 2013; Borde et al. 2014). However information obtained
in this way is also indirect and unreliable because there
are many problems with measurements and especially with
selection and interpretation of weak lines. Moreover
properties of the $Ly-\alpha$ forest are very sensitive to the
spatial variations of the poorly known UV background
(Demia\'nski et al. 2006; Kollmeier et al. 2014)).

 Attempts of direct detection of DM particles by
DAMA (Bernabei 2008, 2010), CRESST-II (Angloher et al. 2012),
and SuperCDMS (Agnese 2013) experiments and others (see
review in Gaitskill 2014) have not yet produced reliable
positive results. Hence up to now we have no reliable
estimates of the mass, the nature and properties of DM
particles. However, the large amount of data already
accumulated by the  LHC could soon lead to detection of
DM particles.

Now one of the most popular candidate for the DM particle is
the sterile neutrino with a mass in the keV range (see, e.g.,
reviews of Feng 2010; Boyarsky et al., 2009c, 2013; Kusenko
2009; Kusenko \& Rosenberg 2013; Dreves 2013; Horiuchi et al.
2013; Marcovi$\breve{c}$ \& Viel 2013; Pontzen \& Governato
2014). Sterile neutrinos can be produced during the inflation
period or later via various processes. In particular a
possible decay of some sort of sterile neutrinos is discussed
(e.g., Ferrer \& Hunter 2013; Bulbul et al. 2014; Boyarsky et
al. 2014). So great diversity of possible properties and
processes of generation of sterile neutrinos -- from
inflation and up to decays at some redshifts -- eliminates
correlations between the masses and velocities of these
particles and increases uncertainties in expectations of
their impact on the power spectrum and in particular on
estimates of the damping scales. Let us note that the model
with unstable DM particles implies also the multicomponent
composition of DM.

It is believed that by modeling the three mentioned above
effects namely, the $Ly-\alpha$ forest, density profile
of DM dominated galaxies, the number of observed satellites
of Milky Way,  and observations of the high-z gamma-ray
bursts (see e.g., reviews Boyarsky at al. 2009; Viel et al.
2013; Marcovi$\breve{c}$ \& Viel 2013; de Souza et al. 2013),
it is possible to solve the problem of sterile neutrino and
in particular to restrict its mass by $m_\nu\geq 10 - 20 keV$.
However these estimates restrict the damping scales and the
shape of power spectrum rather than masses of DM particles
(see, e.g., Tremain \& Gunn 1979; Ruffini et al. 2014).
Recently  X-ray emission with the energy $E\sim 3.5 keV$ was
detected from 73 galaxy clusters what can be interpreted
either as a radiative decay of DM particles or as a
recombination line of Ar (Bulbul et al. 2014; Boyarsky et al.
2014). A spatially extended excess of 1 -- 3 GeV gamma rays
from the Galactic Center could be related to annihilation of
DM particles (Daylan et al. 2014; see also Modak et al. 2013).
This discussion shows that now we do not have any reliable
estimates of properties of DM particles.

The observations of DM dominated halos are very well
complemented by numerical simulations which allow to trace
and investigate the early stages of halos formation, as well
as the process of halos virialization and formation of
their internal structure.  The formation of virialized DM halos
begins as the anisotropic collapse in accordance with the
Zel'dovich theory of gravitational instability (Zel'dovich
1970). During later stages the evolution of such objects
becomes again more complex because it is influenced by
their anisotropic environment (see, e.g., real cluster
representations in Pratt et al. 2009). Moreover all the
time the evolution goes through the process of violent
relaxation and merging what is well reproduced by
simulations.

Analysis of simulated halos can be performed in a wide range
of halo masses and redshifts what allows to improve the
description of properties of relaxed halos of galactic and
cluster scales and to link them with the power spectrum of
initial perturbations. Thus it is established that after a
period of rapid evolution the main characteristics of
majority of the high density virialized DM halos become
frozen and their properties are only weakly changing owing
to the accretion of diffuse matter and/or the evolution of
their baryonic component. The basic properties of the
relaxed DM halos are described in many papers (see, e.g.,
Tasitsiomi et al. 2004; Nagai et al. 2007; Croston et al.
2008; Pratt et al., 2009, 2010; Vikhlinin et al. 2006, 2009;
Arnaud et al. 2010; Klypin et al. 2011; Kravtsov \&
Borgani 2012).

For the WDM model the available simulations (see, e.g.,
Maccio 2012, 2013; Angulo, Hahn, Abel, 2013; Schneider,
Smith \& Reed, 2013; Wang et al. 2013; Libeskind et al.
2013; Marcovi$\breve{c}$ \& Viel 2013; Schultz et al. 2014;
Schneider et al. 2014; Dutton et al. 2014) show that
in accordance with expectations the number of low mass halos
decreases and the central cusp in the density profile is
transformed into the core. For larger halos the standard
density profile is formed again but formation of high density
objects is accompanied by appearance of some unexpected
phenomena. Thus Maccio et al. (2012, 2013), Schneider et al.
(2014) confirm the decrease of matter concentration in
the WDM model in comparison with the CDM model but they
inferred that 'standard' WDM model is not able to reproduce
the density profile of low mass galaxies. This inference is
enhanced in Libeskind et al. (2013) where in contrast with
the CDM model their simulation with 1 keV WDM particles
cannot reproduce the formation of the Local Group. In turn,
Schultz et al. (2014) note that in their simulations with
~3keV WDM particles formation of objects at large redshifts
and reionization are oversuppressed. This means that the
simulations of WDM and especially the multicomponent DM
models require further detailed analysis and it is necessary
to put special attention to reproduce links between the mass
function of halos and the power spectra with the free
streaming cut-off.

Without doubt, one can expect a rapid progress in simulations
of more complex cosmological models. However for
preliminary discussions of such models we can use the semi
analytical description of DM dominated objects proposed in our
previous paper (Demia\'nski \& Doroshkevich 2014). It is based
on the approximate analytical description of the structure of
collapsed halos formed by collisionless DM particles. During
the last fifty years similar models have been considered and
applied to study  various aspects of nonlinear matter evolution
(see, e.g., Peebles 1967; Zel'dovich \& Novikov 1983; Fillmore
and Goldreich 1984; Gurevich \& Zybin 1995; Bryan \& Norman
1998; Lithwick \& Dalal 2011). Of course it ignores many
important features of the process of halos formation and is
based on the assumption that the virialized DM halo is formed
during a short period of the spherical collapse at $z\approx
z_{f}$ and later on its parameters vary slowly owing to the
successive matter accretion (see, e.g., discussion in Bullock
et al. 2001; Diemer et al. 2013).

Of course a spherical model can not adequately describe the
real process of halos formation. However properties of the
steady state virialized DM objects are mainly determined by
the integral characteristics of protoobjects and are only
weakly sensitive to details of their evolution. This is
clearly seen in numerous simulations which show that the
Navarro -- Frenk -- White (NFW) density profile (Navarro
et al. 1995, 1996, 1997) very stably appears in majority of
simulated DM halos.

The same simulations show also that properties of the central
cores of virialized DM halos are established during the early
period of halos formation and later on the slow pseudo--
evolution of cores dominates. This means that properties of
halo cores only weakly depend on the halo periphery and are
determined mainly by their mass and the redshift of formation
(Klypin et al. 2011). Using these results we formulate a rough
two parametric description of all the basic properties of 
virialized DM halos. These two basic parameters are the 
virial mass of halos and the redshift of their formation. 
Of course, they actually
characterize the initial entropy of compressed DM particles
and its growth in the process of violent relaxation of the
compressed DM component. But this approach allows us also to
reveal a close correlation between the central pressure and
density of DM halos and the initial power spectrum of density
perturbations.

Of course this approach is applied for the DM dominated halos
only as the dissipative evolution of the baryonic component
distorts properties of the cores of DM halos. Thus we can use
this model in two ways:
\begin{enumerate}
\item{} We can use the central density and pressure of the DM
component and redshift of the DM halos formation, $z_{f}$, as
a parameter that characterizes the 'frozen' properties of the
central region of DM halos. This redshift of formation
correlates also with the virial mass of halos and with the
initial power spectrum.
\item{}Application of the Press -- Schechter (1974) approach
allows us to link together data obtained for a wide range of
mass of virialized objects and thus to restrict the shape of
the initial power spectrum and some characteristics of the DM
particles.
\end{enumerate}

However potential of this approach should not be overestimated.
As usual we can only determine the probability of object
formation and therefore they have statistical character rather
than strict constrains or predictions. Moreover in our
discussion we use observational data of only limited quality
and representativity. Thus we can use only small number ($\leq
100$) of the observed DM dominated objects, their observed
characteristics -- even so important as the virial mass --
are known only with very limited precision and significant
scatter. None the less the potential of the proposed approach
is significant as it considers objects in a wide range of masses. 
We hope that further accumulation of observational data and their
comparison with the high resolution simulations will allow to
essentially improve presented results.

This paper is organized as follows: In Sec. 2 the basic model,
relations and assumptions of our approach are formulated.
In Sec. 3 a short description of an approximate model of DM
dominated halo is presented. Characteristics of the observed
DM dominated objects -- galaxies and clusters of galaxies --
are described in Secs. 4 and 5, and in Sec. 6 properties of
the central entropy and pressure of DM dominated virialized
objects are discussed. Some possible MDM models are
presented in Sec. 7. Discussion and conclusions can be
found in Sec. 8.

\section{Cosmological models with the CDM, WDM and multi
component DM}

\subsection{Cosmological parameters}

In this paper we consider the spatially flat $\Lambda$
dominated model of the Universe with the Hubble parameter,
$H(z)$, the mean critical density $\langle\rho_{cr}\rangle$,
the mean density of non relativistic matter (dark matter
and baryons), $\langle\rho_m(z)\rangle$, and the mean
density and mean number density of baryons, $\langle\rho_b(z)
\rangle\,\&\,\langle n_b(z)\rangle$, given by Komatsu et al.
(2011), Hinshaw et al. (2013):
\[
H^{2}(z) = H_0^2[\Omega_m(1+z)^3+\Omega_\Lambda],\quad H_0=
100h\,{\rm km/s/Mpc}\,,
\]
\[
\langle\rho_m(z)\rangle =
2.5\cdot 10^{-27}z_{10}^3\Theta_m\frac{g}{cm^3}=
3.4\cdot 10^{4}z_{10}^3\Theta_m\frac{M_\odot}{kpc^3} \,,
\]
\be
\langle\rho_b(z)\rangle = {3H_0^2\over 8\pi G}\Omega_b(1+z)^3
\approx 4\cdot 10^{-28}z_{10}^3\Theta_b\frac{g}{cm^3} \,,
\label{basic}
\ee
\[
\langle\rho_{cr}\rangle=\frac{3H^2}{8\pi G},\quad z_{10}=
\frac{1+z}{10},\quad \Theta_m=\frac{\Omega_mh^2}{0.12},
\quad \Theta_b=\frac{\Omega_bh^2}{0.02}\,.
\]
Here $\Omega_m=0.24\,\&\,\Omega_\Lambda=0.76$ are the mean
dimensionless density of non relativistic matter and dark
energy, $\Omega_b\approx 0.04$ and $h=0.7$ are the
dimensionless mean density of baryons, and the Hubble
constant measured at the present epoch. Cosmological
parameters presented in the recent publication of the Planck
collaboration (Ade et al. 2013) slightly differ from those
used above (\ref{basic}).

For this model the evolution of perturbations can be described
with sufficient precision by the expression
\be
\delta\rho/\rho\propto B(z),\quad B^{-3}(z)\approx \frac{1-
\Omega_m+2.2\Omega_m(1+z)^3}{1+1.2\Omega_m}\,,
\label{Bz}
\ee
(Demia\'nski, Doroshkevich, 1999, 2004, 2014; Demia\'nski et
al. 2011) and for $\Omega_m\approx 0.25$ we get
\be
B^{-1}(z)\approx\frac{1+z}{1.35}[1+1.44/(1+z)^3]^{1/3}\,.
\label{bbz}
\ee
For $z=0$ we have $B=1$ and for $z\geq 1,\, B(z)$ is reproducing
the exact value with accuracy better than 90\%.

For $z\gg 1$ these relations simplify. Thus, for the Hubble
constant and the function $B(z)$ we get
\be
H^{-1}(z)\approx \frac{2.7\cdot
10^{16}}{\sqrt{\Theta_m}}s \left[\frac{10}{1+z}\right]^{3/2},
\quad B(z)\approx \frac{1.35}{1+z}\,.
\label{bzz}
\ee

\subsection{Power spectrum in the WDM models}

The transfer function for the WDM model with
thermalized DM particles was obtained by Bode,
Ostriker and Turok (2001) and more recently in Viel
et al. (2005) (see also Polisensky \& Ricotti 2011;
Marcovi$\breve{c}$  \& Viel 2013). In these papers the
transfer function was written as
\be
T_{WDM}\approx [1+(\alpha_w q)^{2.25}]^{-4.46}\,,
\label{viel}
\ee
\[
q=\frac{k}{\Omega_mh^2}Mpc,\quad
\alpha_w=6\cdot 10^{-3}\left(\frac{\Omega_mh^2}{0.12}\right)^{1.4}
\left(\frac{1 keV}{m_w}\right)^{1.1}\,,
\]
where $k$ is the comoving wave number.

\begin{figure}
\centering
\epsfxsize=6.5cm
\epsfbox{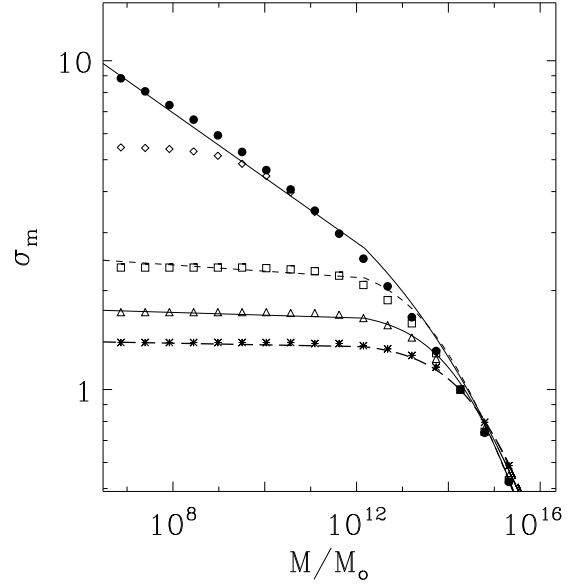}
\vspace{1cm}
\caption{Correlation functions of the matter density $\sigma_m$
are plotted vs. the virial mass of objects, $M_{vir}/M_\odot$,
for the CDM power spectrum (Bardeen et al. 1986) (points) and
for the WDM power spectrum (\ref{viel}) with $m_w=1,\,0.1,
\,0.05,\,\&\,0.03 keV$ (rhombus, squares, triangles and stars).
Fits (\ref{fwdm005}), and (\ref{fcdm}) are plotted by solid
lines, fits (\ref{fwdm01}) and (\ref{fwdm003}) are plotted
by dashed and long dashed lines.
}
\label{crw}
\end{figure}

The correlation function of the density perturbations
\be
\sigma_m^2(R)=4\pi\int_0^\infty p(k)W^2(kR)k^2dk\,,
\label{sig_m}
\ee
with the standard top--hat window function $W(kR)$ has been
discussed already many years ago (see, e.g., Loeb \& Barkana, 
2001). Following the Press -- Schechter approach (Press\,
\&\,Schechter 1974; Peebles 1974; Peacock \& Heavens 1990;
Bond et al. 1991; Mantz et al. 2010) we can link the redshift of
formation $z_f$ of virialized objects with mass $M_{vir}$ with
the correlation function $\sigma_m(M_{vir})$ by the condition
\be
B(z_{f})\,\sigma_m(M_{vir})\approx const\,.
\label{bsig}
\ee
However this approach does not allow us to obtain an independent
estimate of the small scale amplitude of perturbations. More
detailed comparison of the mass dependence of the redshift of
formation of galaxies and clusters of galaxies requires much
more precise estimates of the observational parameters of both
galaxies and clusters of galaxies. In particular thus defined
$\sigma_m$ depends upon the shape of the transfer function
(\ref{viel}) and for the MDM model more accurate determination
of this function is required.

For the transfer function (\ref{viel}) the correlation function
of the density fluctuations is plotted in Fig. \ref{crw} for
four values of $m_w=1,\,0.1, \,0.05,\,\&\,0.03 keV$ and can be
fitted with reasonable precision by the expressions
\be
\sigma_m(M)=2.5/(1+0.12M_{12}^{0.45}),\quad m_w=0.1keV\,,
\label{fwdm01}
\ee
\be
\sigma_m(M)=1.8/(1+0.05M_{12}^{0.45}),\quad m_w=0.05keV\,,
\label{fwdm005}
\ee
\be
\sigma_m(M)=1.4/(1+0.03M_{12}^{0.45}),\quad m_w=0.03keV\,.
\label{fwdm003}
\ee
For comparison in the same Figure the correlation function for
the standard CDM model  is also plotted. It is well fitted by
the expression (Klypin et al., 2011):
\be
\sigma_m(M)=3.9M_{12}^{-0.077}/(1+0.18M_{12}^{0.133}+
0.14M_{12}^{0.333})\,.
\label{fcdm}
\ee
As is seen from Fig. \ref{crw} for the mass of the WDM particles
$m_w\geq 1keV$ functions $\sigma_m$ differ from the CDM one
for objects with moderate mass $M_{vir}\leq 10^9M_\odot$ only.
For the WDM model with less massive particles $1keV\geq
m_w\geq 0.03keV$ the damping mass
falls in the range $10^{10}\leq M_{vir}/M_\odot\leq 10^{14}$.
Such damping strongly decelerates formation of objects with
mass $M_{obj}\leq M_{dmp}$ (see, e.g., Schultz et al. 2014).

\subsection{Power spectrum in  MDM models}

 The evolution of perturbations in MDM models was
discussed many times in different approximations (see e.g.
Grishchuk \& Zel'dovich 1981; Turner et al. 1984; Doroshkevich
et al. 1984; Boyarsky et al. 2009b; Anderhalden et al. 2012).

It is important that in contrast with the CDM or WDM models
with one mass of particles in MDM models
the shape of transfer function is time dependent. Indeed
even relatively small fraction of  WDM particles with $m_w\leq
m_{cdm}$ significantly decelerates the growth of CDM perturbations
at small scales as compared with the standard CDM model. This
deceleration decreases with time the height of plateau in the
transfer function of MDM model for larger $k$ what leads to
the progressive damping of the power spectrum at such scales.
This problem was briefly discussed by Boyarsky et al. (2009b)
where the contribution of the CDM power spectrum, $g_{cdm}$,
to the full power spectrum at redshifts $z\leq 1$ is roughly 
linked with the fraction of CDM matter $f_{cdm}$ as
\be
g_{cdm}\approx f_{cdm} 10^{2.58(1-f_{cdm})}\,.
\label{decay}
\ee

This fit neglects the time dependence of the height of plateau
in the MDM transfer function what misrepresents
the shape of the MDM power spectrum and the function
$\sigma_m(M_{vir})$ and thus increases uncertainties of
our consideration. Non the less allowing for qualitative
character of our approach, further on, we will use this
relation to roughly link the spectral and matter
fractions, $g_{cdm}$ and $f_{cdm}$. Unfortunately we
do not see a simple way to improve on these disadvantages and
in what follows we will determine the MDM
correlation function by the relation
\be
\sigma_m(M)=\sqrt{g_{cdm}\sigma^2_{cdm}+
g_{wdm}\sigma^2_{wdm}}\,.
\label{fmdm}
\ee
This relation implies statistical independence of perturbations 
in the CDM and WDM mediums what is only approximately correct.
More precise conclusions can be obtained with high resolution
numerical simulations.

\section{Physical model of halos formation}

Properties of both simulated and observed virialized
objects -- galaxies and clusters of galaxies -- are usually
described in the framework of spherical models such as Navarro
-- Frenk -- White (NFW) (Navarro et al. 1995, 1996, 1997;
Ludlow et al. 2013), Burkert (1995) or isothermal models.
In this paper we link the virial mass of DM halos $M_{vir}$
with the redshift of their formation, $z_f$. For this purpose we
use the  spherical model of  DM halos formation which
was discussed in many papers (see, e.g., Peebles 1967,
Umemura et al., 1993, Bryan \& Norman 1998).  Here we
will briefly describe the main properties of this model.

It is commonly accepted that in the course of complex nonlinear
condensation the DM forms stable virialized halos with a more or
less standard density profile. Numerical simulations show that
the virialized DM halos are formed from initial perturbations
after a short period of rapid complex evolution. For example
such virialized objects are observed as isolated galaxies and/or
as high density galaxies embedded within clusters of galaxies,
filaments, superclusters or other elements of the Large
Scale Structure of the Universe. This approach also allows us
to estimate the redshift when the observed DM dominated
objects such us the dSph galaxies and clusters
of galaxies were formed. Of course, this model ignores
all details of the complex process of halos formation. But it
allows to obtain a very simple, though rough, description of
this process
and introduces some order of objects formation.

Our simple physical model of halos formation is
based on the following assumptions:
\begin{enumerate}
\item{} We assume that at redshift $z=z_{f}$ the evolution
of DM perturbations results in the formation of spherical
virialized DM halos with masses $M_{vir}=M_{13}\cdot 10^{13}
M_\odot$ and the central densities $\rho_c$.
\item{} We do not discuss the dynamics of DM halos
evolution which is accompanied by the progressive matter
accretion, the growth of the halos masses and corresponding
variations of other halos parameters. The real process of halos
formation is extended in time what causes some ambiguity in
their parameters such as the halos masses and the redshift of
their formation (see, e.g. discussion in Diemand, Kuhlen \&
Madau 2007; Kravtsov \& Borgani 2012). In the proposed model
the redshift of halo formation, $z_{f}$, is identified with
the redshift of collapse of the homogeneous spherical cloud
with the virial mass $M_{vir}$,
\be
1+z_{f}\approx 0.63 (1+z_{tr})\,,
\label{zform}
\ee
where $z_{tr}$ is the redshift corresponding to the turn
around moment of the dust cloud evolution (see discussion
of spherical model in Umemura, Loeb\,\&\,Turner 1993).
\item{}We assume that in the course of DM halo formation
the main fraction of the baryonic component is heated by the
accompanied shock waves up to the temperature and pressure
comparable with the virialized values of the DM component.
These processes are responsible for the formation of equilibrium
distribution of the baryonic component.
\item{} We assume that some (random) fraction of the
compressed baryons is collected into a system of subclouds
which are rapidly cooled and transformed into high density
subclouds. Thus, the virialized halo configuration is composed
of the DM particles, the hot low density baryonic gas, and
cold high density baryonic subclouds.
\item{} Transformation of less massive DM halos into the
first observed galaxies with some fraction of stars was discussed
in (Demia\'nski \& Doroshkevich 2014).
\end{enumerate}

The evolution of the cooled subclouds  can be very
complex. It can be approximated by the isobaric mode of the
thermal instability and therefore it does not preserve the
compact shape of the cooled subclouds. As was discussed in
Doroshkevich and Zel'dovich (1981) the motion of such subclouds
within the hot gas leads to their deformation and less massive
subclouds could be disrupted and even dissipated. The complex
aspherical shape of such subclouds makes their survival
problematic and requires very detailed investigation to
estimate their evolution even in simulations. These problems
are however beyond the scope of this paper.

The basic parameters of the discussed model --  the virial
mass, $M_{vir}$, central density, $\rho_c$, and concentration,
$C$, connect the central and mean densities of objects
by expressions
\be
M_{vir}=4\pi/3R_{vir}^3\Delta_v\langle\rho_{cr}(z_{f})
\rangle =4\pi \rho_cR_{vir}^3C^{-3}f_m(C)\,,
\label{mvirial}
\ee
\be
\frac{C^3(z_{f},M_{vir})}{f_m(C)}=\frac{3\rho_c}{\Delta_v
\langle\rho_{cr}(z_{f})\rangle},\quad f_m=\int_0^C dx\,x^2
\frac{\rho(x)}{\rho_c}\,,
\label{nfw-vir}
\ee
\[
C=R_{vir}/r_c,\quad x=r/r_c,\quad \Delta_v=18\pi^2\approx 200\,.
\]
Here $R_{vir}$ and $r_c$ are the virial radius of halo and the
radius of central core (for the NFW or Burkert (1995) models)
or size of the isothermal core, $\rho_c$ and $\langle\rho_{cr}
(z_{f})\rangle$ are the central density of halo and the 
critical density of the Universe at redshift $z_f$ (\ref{basic}). 
The value of mean overdensity, $\Delta_v$, was derived from 
the simple model of spherical collapse that ignores the 
influence of complex anisotropic
halos environment (see, e.g., Bryan \& Norman 1998; Vikhlinin
et al. 2009; Lloyd--Davies et al. 2011). The well known factor
$f_m(C)\sim 1$ links the virial mass of an object with its
concentration. Impact of the factor $f(C)$ can be determined
by the method of successive approximations using, for example,
the NFW density profile with
\[
f_m(C)=\ln(1+C)-C/(1+C)\,,
\]
or quite similar expression for the Burkert (1995) model.

Application of this approach is possible for a given
concentration $C(z_f,M_{vir})$. Below in section 4 we will
use suitable fits for $C(z_f,M_{vir})$ given by Klypin et al.
(2011).

Of course, this approach  provides the qualitative description
only and has limited predictive power.
Thus, it ignores the complex anisotropic successive matter
compression within filaments and walls before formation of
compact halos, it ignores the effects produced by mergers,
by anisotropic halo environment and so on. More detailed
description can be achieved in the framework of more complex
aspherical model which would take into account possible
impact of such ignored effects as a random scatter of redshift
of halos formation and other halo characteristics for a
given virial mass. However use of such more complex
models is not verified and direct comparison of observed
and simulated objects seems to be more successful.

In central regions of halos the gas pressure is supported by
adiabatic inflow of high entropy gas from outer regions of halos
what leads to progressive concentration of baryonic component
within central regions of halos and to formation of massive
baryonic cores (see, e.g. Wise\,\&\,Abel 2008; Pratt et al.
2009; McDonald et al. 2013).

\section{Observed characteristics of clusters of galaxies}

For our analysis we use more or less reliable observational 
data for $\sim 150$ DM dominated clusters of galaxies, for 
$\sim$40 dSph, and $\sim$20 other DM dominated galaxies. 
The analysis of observations of the dSph galaxies in the
framework of accepted model was performed in Demia\'nski 
\& Doroshkevich (2014). Here we continue this analysis and
compare observational results with theoretical expectations.

\subsection{Redshift of formation of clusters of galaxies}

Now there are more or less reliable observational data at
least for $\sim 300$ clusters of galaxies (Pointecouteau et
al. 2005; Arnaud et al., 2005; Pratt et al., 2006; Zhang et
al., 2006; Branchesi et al., 2007; Vikhlinin et al., 2009;
Pratt et al. 2010; Suhada et al. 2012; Moughan et al. 2012;
Fo\H{e}x et al. 2013; Bhattacharya et al. 2013). However,
the central cluster characteristics are not directly observed
and are obtained by a rather complex procedure (see, e.g.,
Bryan\,\&\,Norman 1998; Vikhlinin et al. 2009; Lloyd--Davies
et al. 2011; McDonald et al., 2013).

In spite of the speedy progress in investigations of the
clusters of galaxies recent publications discuss mainly the
observations of general cluster characteristics such as their
redshift $z_{obs}$, virial mass, $M_{vir}$, radius, $R_{vir}$,
and average temperature, $T_x$. It is important that only
the observed redshift of clusters, and their averaged
temperature are really measured while other characteristics
are usually derived using empirical correlations (see, e.g.,
Pointecouteau et al. 2005; Vikhlinin et al., 2006; Nulsen,
Powell, \& Vikhlinin, 2010). It is interesting that some of these
correlations can be expressed in the standard form used for
description of virialized objects,
\[
\beta=\frac{U}{W}\approx \frac{GM_{vir}}{R_{vir}T_{x}}=const\,,
\]
where $U\,\&\,W$ are the gravitational and internal energy of
the object. Thus for 179 observed clusters with masses $2\leq
M_{13}=M_{vir}/10^{13}M_\odot\leq 250$ we have
\be
\langle\beta\rangle=\left\langle\frac{M_{13}}{R_{Mpc}T_{kev}}
\right\rangle\approx 7.16(1\pm 0.08)\,.
\label{virial}
\ee
In spite of limited precision of these determinations small
scatter of $\beta$ demonstrates both stability of observational
methods and similarity of internal structure of clusters.

However in this paper we are mainly interested in discussion
of more stable central regions of clusters and, in particular,
in concentrations and in the central pressure of clusters.
Unfortunately the body of such data is very limited.

In this section we consider properties of the central regions
of virialized DM halos using the approximation presented
in Klypin et al. (2011), Prada et al. (2012), Angulo et al.
(2012). We characterize halos by their virial mass $M_{vir}$
and redshift of formation, $z=z_{f}$. We assume that at
$z\leq z_{f}$ the halos mass and mean temperature and density
remain almost the same and as usual we take
\be
\langle\rho_{cl}(z_{f})\rangle\simeq 500\rho_m(z_{f})\approx
1.25\cdot 10^{-27}(1+z_{f})^3 g/cm^3\,.
\label{rho_mns}
\ee
The mean baryonic number density of relaxed halos is
\[
\langle n_b(z_{f})\rangle=1.5\cdot 10^{-4}(1+z_{f})^3
\Theta_bcm^{-3}\,.
\]

\subsection{Observed characteristics of clusters
formed at small redshifts}

For clusters of galaxies with mass $M_{vir}= M_{13}\cdot
10^{13}M_\odot$ formed at redshifts $z\leq 1$ the
concentration $C(M_{vir},z_{f})$ is given by the
expression (Klypin et al. 2011)
\be
C(M_{vir},z_{f})\approx 7.5B^{4/3}(z_{f})M_{13}^{-0.09}\,,
\label{clg}
\ee
and for such clusters determination of the redshift $z_{f}$ is
difficult. Indeed, comparison of the central density
$\rho_c(M_{vir}, z_{f})$ with the mean density shows that,
\[
\rho_c(M,z_f)\approx\frac{\langle\rho_{cl}\rangle C^3}{3f_m(C)}
\approx 1.8\cdot 10^{-25}\frac{g}{cm^3}
\frac{D^3(z_{f})}{M_{13}^{0.27}f_m(C)}\,\,,
\]
\be
n_c(M,z_{f})\approx 0.1M_{13}^{-0.27}cm^{-3}D^3(z_f)/f_m(C),
\label{dnfw}
\ee
\[
D(z_{f})=(1+z_{f})B^{4/3}(z_{f})\,,
\]
and the function $f_m(C)\sim 1$ was introduced by
(\ref{nfw-vir}). It is important that $D(z)$ only very
weakly depends on the redshift, $D(z)\sim 1.1$ for
$0\leq z\leq 1$ and the precision of available data set
makes it difficult to reveal evolution of clusters and
to use them for discussion of cosmological problems.

For 18 nearby clusters observed at $\langle z_{obs}\rangle=
0.095$ and with masses $8\leq M_{13}\leq 120$ (Pointecouteau
2005; Vikhlinin 2006; Bhattacharya et al. 2013) the measured
values of concentration are
\[
\langle C\rangle=3.46(1\pm 0.15),\quad f_m(C)\approx 0.73\,.
\]
These data allow us to roughly estimate  the central pressure,
$P_c$, baryon number density, $n_c$, and entropy, $S_c$
in clusters and the redshift of their formation as
\be
\langle P_c\rangle\approx 23.1(1\pm 0.5)eV/cm^3\,,
\label{p18}
\ee
\[
\langle n_c\rangle\approx 0.5\cdot 10^{-2}(1\pm 0.5)cm^{-3},
\quad \left\langle\frac{1+z_{f}}{1+z_{obs}}\right\rangle
\approx 1.66\,,
\]
\be
\langle S_b\rangle=\langle P_c/n_c^{5/3}\rangle\approx 180(1
\pm 0.7) cm^2keV\,,
\label{s18}
\ee
\be
\langle 1+z_{f}\rangle\approx 1.8(1\pm 0.2),\quad
\langle B^{-1}(z_{f})\rangle\approx 1.5(1\pm 0.14)\,.
\label{zcr1}
\ee
These results confirm that clusters are formed earlier than
they are observed and it is necessary to bear in mind this
difference in the course of interpretation of cluster parameters.

\subsection{Properties of the DM halos formed
at larger redshifts}

\begin{figure}
\centering
\epsfxsize=6.5cm
\epsfbox{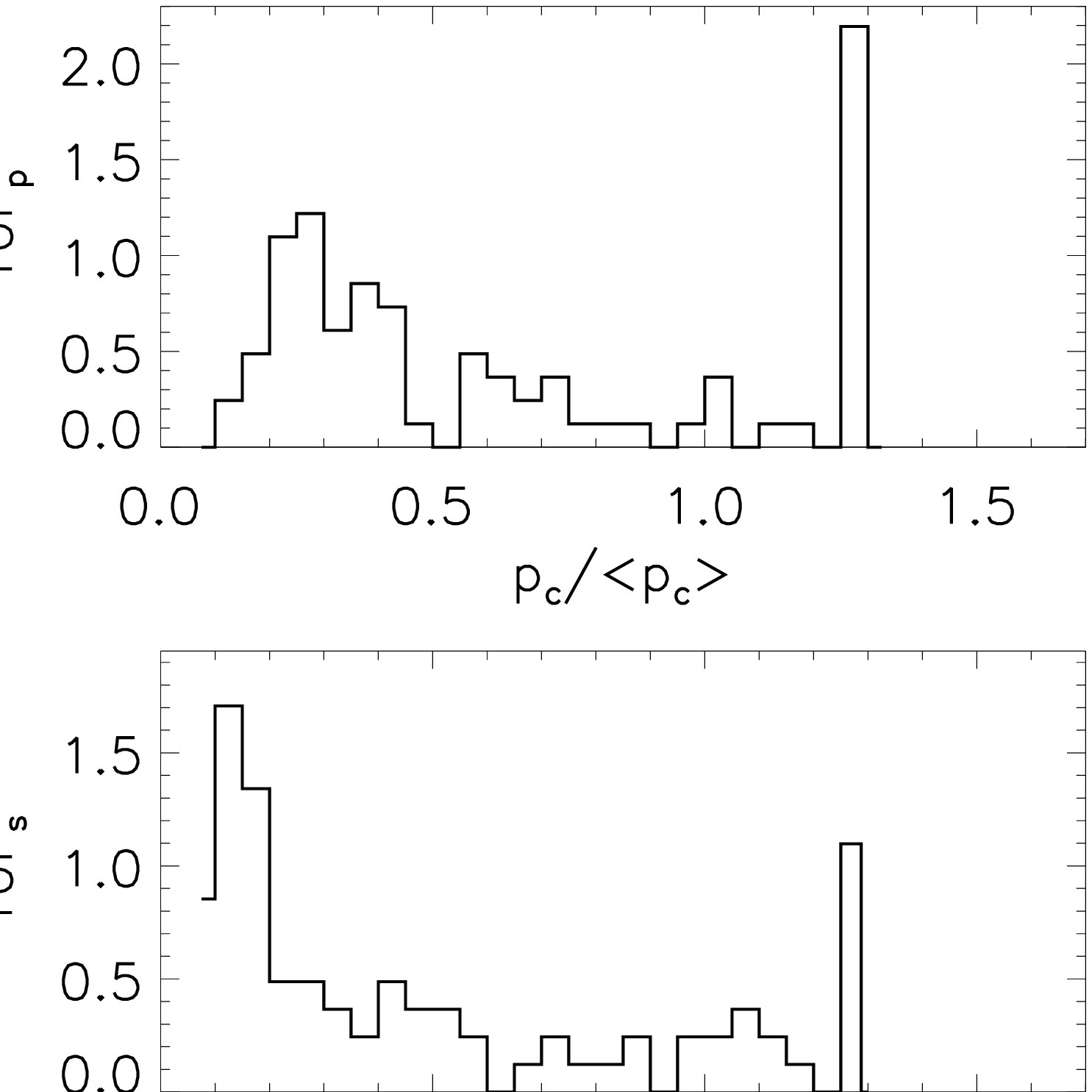}
\vspace{1cm}
\caption{For 83 clusters from the SPT--sample the distribution
functions of the central pressure, $F_p(p_c/\langle
p_c\rangle)$ and central entropy $F_s(0.5s_c/\langle s_c
\rangle)$ are plotted.
}
\label{hst-83}
\end{figure}

\subsubsection{Theoretical expectations}

For description of galactic scale halos or earlier formed DM
halos of cluster scale it is convenient to use other
approximation of halo parameters (Klypin et al. 2011;
Demia\'nski \& Doroshkevich, 2014)
\be
C\approx 0.18M_{13}^{1/6}(1+z_{f})^{7/3}
=0.18M_{13}^{-1/15}\eta^{7/3}\,,
\label{nfw-c}
\ee
\[
\eta=(1+z_{f})M_{13}^{0.1}\,.
\]
Comparison with observations shows that for the DM dominated 
relaxed objects 
\be 
\eta\sim 3.2 - 3.3\,, 
\label{eta} 
\ee
remains almost the same in a wide range of virial masses and 
redshifts $z_{f}$.   For such halos using (\ref{nfw-vir}) and 
(\ref{nfw-c}) we expect to have for the central density of the 
DM matter 
\be 
\rho_c(M,z)\approx
\rho_0(1+z_{f})^{10}M_{13}^{1/2} =\rho_0\eta^{10}M_{13}^{-1/2}\,, 
\label{dnfw} 
\ee
\[
\rho_0=
1.1\cdot 10^{-8}\Theta_\rho \frac{M_\odot}{pc^3},\quad
\Theta_\rho=\frac{\delta_r}{f_m(C)}\frac{\Delta_v}{200}
\Theta_m\,,
\]
and for the central baryonic number density
\be
n_{b}=0.14 cm^{-3}(\eta/3.3)^{10}M_{13}^{-1/2}
\Theta_b\Theta_\rho\,.
\label{nnc}
\ee
Here $\Theta_\rho$ describes the random variations of the
central density ($\delta_r$) and uncertainties in determination
of the observed parameters. As was shown in Demia\'nski \&
Doroshkevich (2014) for the concentration (\ref{nfw-c})
the central pressure depends only upon $\eta(M_{vir},z_{f})$
(\ref{eta}) and we get
\be
P_c(M_{vir},z_{f})\approx P_0(\eta/3.3)^{40/3},\quad
P_0\approx 28eV/cm^3\,.
\label{pcc}
\ee
It can be expected that it is only weakly sensitive to
other parameters of clusters. For the central entropy we get
\be
S_c=P_c/n_b^{5/3}\approx 0.76(3.3/\eta)^{10/3}M_{13}^{5/6}
cm^2keV\,.
\label{scc}
\ee

\subsubsection{Properties of SPT clusters}

These expectations can be compared with parameters of 83 
clusters selected by the South Pole Telescope (Reinhardt et
al. 2013; McDonald et al. 2013; Ruel et al. 2013; Saliwanchik 
et al. 2013). For this SPT-sample of clusters the central
baryonic density, temperature and entropy are given at 
radius $r\leq 0.012R_{500}$ where $R_{500}$ is the radius 
of a sphere within which the average density is 
500$\varrho_{crit}(z_{abs})$ . For 31 clusters also standard 
X-ray masses derived from Chandra observations (Ruel et 
al. 2013) are known.

For this sample the distribution functions of the central
pressure and entropy are plotted in Fig. \ref{hst-83} where
\be
\langle P_c\rangle\approx 146eV/cm^3,\quad
\langle S_c\rangle\approx 145keV cm^2\,.
\label{spt83}
\ee

As is seen from this Figure this sample is clearly divided
into two groups. One of them contains 39 clusters with higher
central pressure and low entropy
\be
\langle P_{col}\rangle=270eV/cm^3,\quad \langle S_{col}\rangle=
100keV cm^2\,.
\label{cold-cl}
\ee
For these clusters both the baryonic density and pressure
are extremely high, while the entropy is small. This is
explained by strong cooling and clumping of the observed
gaseous component. It can be expected that in these clusters
owing to the thermal instability the baryonic matter forms
two fractions, one of which is represented by a system of high
density low temperature clouds and the other is composed of
high temperature low density gas. In this case the measured
density relates to the denser fraction while the temperature
relates to the hot gas and random velocities of clouds
(see, e.g., Khedekar et al. 2013). If this interpretation is
correct then the measured $P_c$ and $S_c$ (\ref{cold-cl})
are artificial and the real central pressure and entropy of
the hot component are close to that measured for the other 44 
clusters presented below while the entropy of the cold 
component is less than (\ref{cold-cl}).

For the subsample of 44 clusters with $1\leq M_{13}
\leq 100$ and moderate central pressure $P_c\leq 70 eV/cm^3$,
we have 
\be
\langle P_c\rangle\approx 36.1(1\pm 0.37)eV/cm^3,\quad
\eta\approx 3.36(1\pm 0.04)\,,
\label{pc0}
\ee
\[
\kappa(n_c,T_c)\approx -0.67\,,
\]
where $\kappa(f_1,f_2)$ is the standard correlation coefficient
\be
\kappa(f_1,f_2)=(\langle f_1f_2\rangle-\langle f_1\rangle
\langle f_2\rangle)/\sigma_1/\sigma_2\,.
\label{kappa}
\ee
However this subsample is naturally divided into two groups. 
Hotter group accumulates 24 clusters with
\[
\langle n_c\rangle = 0.5\cdot 10^{-2}(1\pm 0.4)cm^{-3},
\quad\kappa(n_c,T_c)\approx -0.65,
\]
\be
\langle P_c\rangle\approx 36.2(1\pm 0.35)eV/cm^3,\quad
\langle T_c\rangle = 7.8(1\pm 0.3)keV\,,
\label{hot}
\ee
\[
\langle S_b\rangle = 305(1\pm 0.5)cm^2keV,
\quad\kappa(P_c,S_b) = -0.2\,.
\]
For 20 colder clusters we get
\[
\langle n_c\rangle = 1.8\cdot 10^{-2}(1\pm 0.6)cm^{-3},\quad
\kappa(n_c,T_c)\approx -0.54\,,
\]
\be
\langle P_c\rangle\approx 35.9(1\pm 0.39)eV/cm^3,\quad
\langle T_c\rangle = 2.3(1\pm 0.4)keV\,,
\label{cold}
\ee
\[
\langle S_b\rangle = 41(1\pm 0.6)cm^2keV,\quad
\kappa(P_c,S_b) = -0.1\,.
\]
The noticeable correlation of the cluster temperature and
density together with negligible correlation between cluster
pressure and entropy allows to obtain more detailed description
of the process of DM compression and successive violent
relaxation of the compressed matter. As is seen from (\ref{hot},
\,\ref{cold}) both the central pressure and entropy contain
some regular term depending upon characteristics of clusters
(for (\ref{hot})) and upon cooling of baryonic component (for
(\ref{cold})). The random fluctuations of the pressure and
entropy, $\delta P_c/P_c\,\&\,\delta S_c/S_c$, with
\[
\sigma_p^2=\langle(\delta P_c/P_c)^2\rangle\approx 0.16,\quad
\sigma_s^2= \langle(\delta S_c/S_c)^2\rangle\approx 0.3\,,
\]
are (almost) independent. Neglecting their correlation,
$\kappa(P_c,S_b)=0$, it is easy to see that the dispersions
of the central density, $\sigma_n$, and temperature,
$\sigma_T$, and their correlation coefficient, $\kappa(n_c,
T_c)$, are simple functions of $\sigma_p\,\&\,\sigma_s$.
Indeed, for $P_c=S_bn_c^{5/3}=T_cn_c$
\[
\sigma_n=0.6\sqrt{\sigma_p^2+\sigma_s^2}\sim 0.45,\quad
\sigma_T=0.6\sqrt{0.44\sigma_p^2+\sigma_s^2}\sim 0.36,
\]
\[
\kappa(n_c,T_c)\approx -0.6,\, {\rm for}\,\,
\sigma_p/\sigma_s\approx 0.7\,,
\]
what is comparable with (\ref{hot},\,\ref{cold}).
These properties of perturbations seem to suggest that
uncorrelated pressure and entropy perturbations are fundamental
while perturbations of density and temperature can be
considered as their consequence.

The differences in properties of 24 hotter and 20 colder
clusters could be mainly caused by cooling of the baryonic
component. Indeed, the isobaric thermal instability results
in formation of a two phase medium -- cold denser clouds
moving within hot gas -- but it does not perturb the gas
pressure. Weak scatter of the central pressure of these
clusters (\ref{pc0}) shows that this pressure as well as
the entropy of 24 hot clusters characterize the regular
process of violent relaxation of the dominant DM component
while the decrease of entropy for 20 colder clusters is
naturally explained by the cooling of baryonic component.
The random uncorrelated scatter of the central pressure and
entropy are naturally related to random uncorrelated
variations in the initial state of the compressed matter.

\begin{figure}
\centering
\epsfxsize=6.5cm
\epsfbox{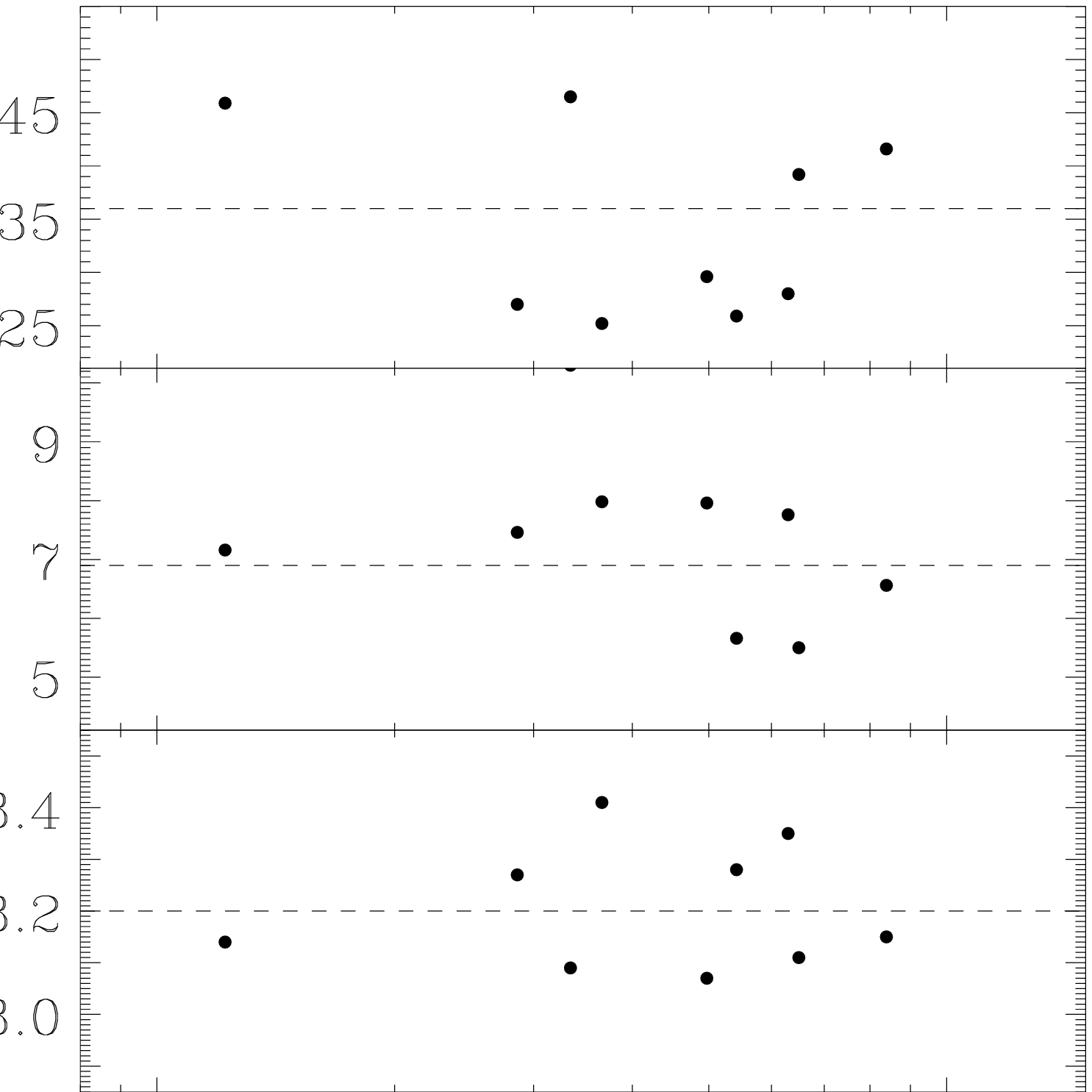}
\vspace{1cm}
\caption{For 9 clusters from the SPT--sample the virial
factor $\beta = M_{13}/R_{vir}/T$, redshift creation,
$1+z_{f}$ and central pressure, $P_c eV/cm^3$ are plotted
vs. the masses $M_{14}$. Fits (\ref{cls09},\,\ref{pc1},\&\,
\ref{zcr2}) are plotted by dashed lines
}
\label{cls-83}
\end{figure}

For 9 clusters of this subsample also their X-ray mass was
determined. For these clusters with $10\leq M_{13}\leq 80$
and
\be
\langle\beta\rangle=\left\langle\frac{M_{13}}{R_{x}T_x}
\right\rangle\approx 8(1\pm 0.18),\quad
\langle 1+z_{obs}\rangle = 1.7(1\pm 0.12)\,,
\label{cls09}
\ee
we get for the pressure, $P_c$ \& $\eta$, baryonic number
density, $n_b$, and the entropy, $S_b$:
\[
\langle n_b\rangle\approx 0.7\cdot 10^{-2}(1\pm 0.4)cm^{-3}\,,
\]
\be
\langle P_c\rangle\approx 34.2(1\pm 0.24)eV/cm^3\,,
\label{pc1}
\ee
\[
\langle\eta\rangle\approx 3.35(1\pm 0.02)\,,
\]
\[
\langle S_b\rangle = 200(1\pm 0.7)cm^2keV\,.
\]

It is important that if the central pressures of both 
subsamples (\ref{hot}) and (\ref{cold}) are close to the
expected one (\ref{pcc}) then the observed central entropies 
(\ref{hot}) and (\ref{cold}) noticeably exceed the
expected value (\ref{scc}). Reasons of these divergences 
are unknown but it can be expected that the progressive growth
of small scale perturbations and their following randomization 
can sufficiently increase the large scale entropy of
matter compressed in clusters of galaxies. This problem 
requires further analysis in simulations.

The redshift of cluster formation, $z_f$ can be determined 
by two methods. Firstly, we could use the measurements of
the baryonic number density and find the concentration, $C$, 
and $z_f$ from (\ref{nfw-vir}) and (\ref{dnfw}). However,
the $n_c\,\&\,T_c$ are observed at the radius $r\sim 
0.012R_{vir}$, where radial variations of density and 
temperature can be significant, what generates additional 
uncertainties in the estimated value of $z_f$. One can also 
use measurements of pressure and parameter $\eta$ from 
(\ref{pcc}). In central regions of DM halos radial variations 
of pressure are small, scatter of $\eta$ is negligible and 
therefore precision of so determined $z_f$ depends mainly on
the precision of measurements of the cluster mass, $M_{13}$, 
in Eq. (\ref{eta}).

Using the second method we get for these 9 clusters
\be
\langle 1+z_{f}\rangle\approx 2.35(1\pm 0.1)\approx 3.3
(1\pm 0.02)M_{13}^{-0.1}\,,
\label{zcr2}
\ee
\[
\langle B^{-1}(z_{f})\rangle\approx 1.74(1\pm 0.1)\,.
\]

To improve representativity of the subsample of 9 SPT clusters
(\ref{pc1}, \ref{zcr2}) we can extend it up to 31 objects
setting $\eta=3.3$ for all objects with measured X-ray
masses. This assumption agrees with inference (\ref{pc0})
that for all clusters the pressure in central regions is almost
the same and it weakly depends upon the virial masses and
redshifts of formation of clusters. Thus for subsample of 31
clusters with $1\leq M_{13}\leq 150$ we get 
\[
\langle 1+z_{obs}\rangle = 1.66(1\pm 0.15)\,,
\]
\be
\langle 1+z_{f}\rangle=\langle 3.3M_{13}^{-0.1}\rangle
\approx 2.1(1\pm 0.05)\,,
\label{zcr3}
\ee
\[
\langle B^{-1}(z_{f})\rangle = 1.66(1\pm 0.1)\,.
\]
Similarity of these results and those obtained above for the
subsample of 9 clusters (\ref{zcr2}) confirms validity of this
approach. 

\subsubsection{Properties of REXCESS and Bolocam clusters}

It is interesting to compare the pressure, density and redshift
of formation, (\ref{pc1} \& \ref{zcr2}) with the same values
obtained for 9 clusters of REXCESS survey (Arnaud et al.
2010; Pratt et al. 2010) with $10\leq M_{13}\leq 75$
\be
\langle P_c\rangle\approx 21.2(1\pm 0.51)eV/cm^3\,,
\label{pc2}
\ee
\[
\langle n_b\rangle\approx 0.33\cdot 10^{-2}(1\pm 0.54)cm^{-3},
\quad \langle\eta\rangle\approx 3.2(1\pm 0.05)\,,
\]
\be 
\langle 1+z_f\rangle\approx 2.3(1\pm 0.1),\quad 
\langle B^{-1}(z_f)\rangle\approx 1.7(1\pm 0.1)\,. 
\label{zc2} 
\ee
Moderate differences of the pressure $\langle P_c\rangle$ 
observed in very different clusters (\ref{p18},\,\ref{pc0},
\,\ref{pc1},\,\ref{pc2}) demonstrate the high stability of 
these parameters. Unfortunately, for these clusters the 
central temperature was not measured so we have to use our 
estimates (\ref{dnfw}, \ref{pcc}) instead.

This comparison can be continued for the sample of 45 
massive galaxy clusters imaged using the Bolocam for which 
pressure profiles were measured (Sayers et al. 2013). 
These clusters with masses $23\leq M_{13}\leq 420$, 
temperature $4.4keV\leq \langle kT\rangle\leq 14 keV$ and 
outer pressure $2.8eV/cm^3\leq P_{500}\leq 14.9eV/cm^3$ 
are situated at redshifts $0.151\leq z\leq 0.888$. For this 
sample the central pressure at $r\sim 0.07R_{500}$ is 
\be
\langle P_c\rangle\sim 50(1\pm 0.2)eV/cm^3M_{15}^{2/3}
E^{4/3}(z)\,,
\label{boxsz}
\ee
with $E(z)=(1+(1+z)^3/3)$. For clusters with the mass $M_{15}
\leq 1$ this result is also close to (\ref{pc0}), (\ref{pc1}),
and (\ref{pc2}).  For these clusters the central temperature is
also unknown and the central pressure is estimated with
(\ref{dnfw}, \ref{pcc}).

\subsection{Observed density profile of SPT clusters of galaxies}

For all 83 objects of the SPT--catalogs the baryonic density 
slope $\alpha$ at a distance $r\simeq 0.04R_{500}$ is also 
measured and it is justly linked with the process of cooling 
of the compressed gas. Here we confirm that this slope 
strongly correlates with the density of central regions of 
clusters,
\be
\kappa(\alpha, n_b)\approx 0.76\,,
\label{alp}
\ee
where the correlation coefficient $\kappa$ is obtained
according to (\ref{kappa}). So strong correlation indicates
that the steep profile is caused by the limited resolution of
observations. Indeed the cold high density gaseous clouds
naturally arise in the central regions of many clusters owing
to significant density fluctuations that are enhanced by
isobaric modes of the thermal instability. As was demonstrated
in Doroshkevich, Zel'dovich (1981) peculiar motions of such
clouds sometimes lead to their deformation and even complete
disruption.

These results are fully consistent with conclusions of Arnaud 
et al. (2010) where similar entropy and density gradients were
found for the set of REXCESS clusters. The close link between
the power index and the baryonic density is indicated by their
correlation coefficient.

\section{Observed properties of the DM dominated galaxies}

Several DM dominated objects of galactic scale are known.
These are the 41 dSph galaxies, 14 THING
and 10 LSB galaxies. Analysis of these objects can be
performed in the same manner as done above.

\subsection{Observed properties of the dSph galaxies}

During last years properties of dSph galaxies were discussed
in detail in many papers. Thus, the main observed parameters
of 28 dSph galaxies are listed and discussed in Walker et al.
(2009, 2011), Penarrubia et al., (2010), and 13 And galaxies
with similar properties are listed in Tollerud et al. (2012).
These samples include objects in a wide range of masses, $0.1
\leq M_6= M_{gal}/10^6M_\odot\leq 100$, what allows us to reveal
more reliably the mass dependence of their redshift of formation
(Demia\'nski \,\&\,Doroshkevich 2014). In this case we have to
deal with parameters of the central regions at the projected
half--light radius but their reliability is limited and scatter
is large. In spite of this it is interesting to compare
characteristics of these galaxies with characteristics of
clusters of galaxies presented in this Section and with
theoretical expectations.

\begin{figure}
\centering
\epsfxsize=6.5cm
\epsfbox{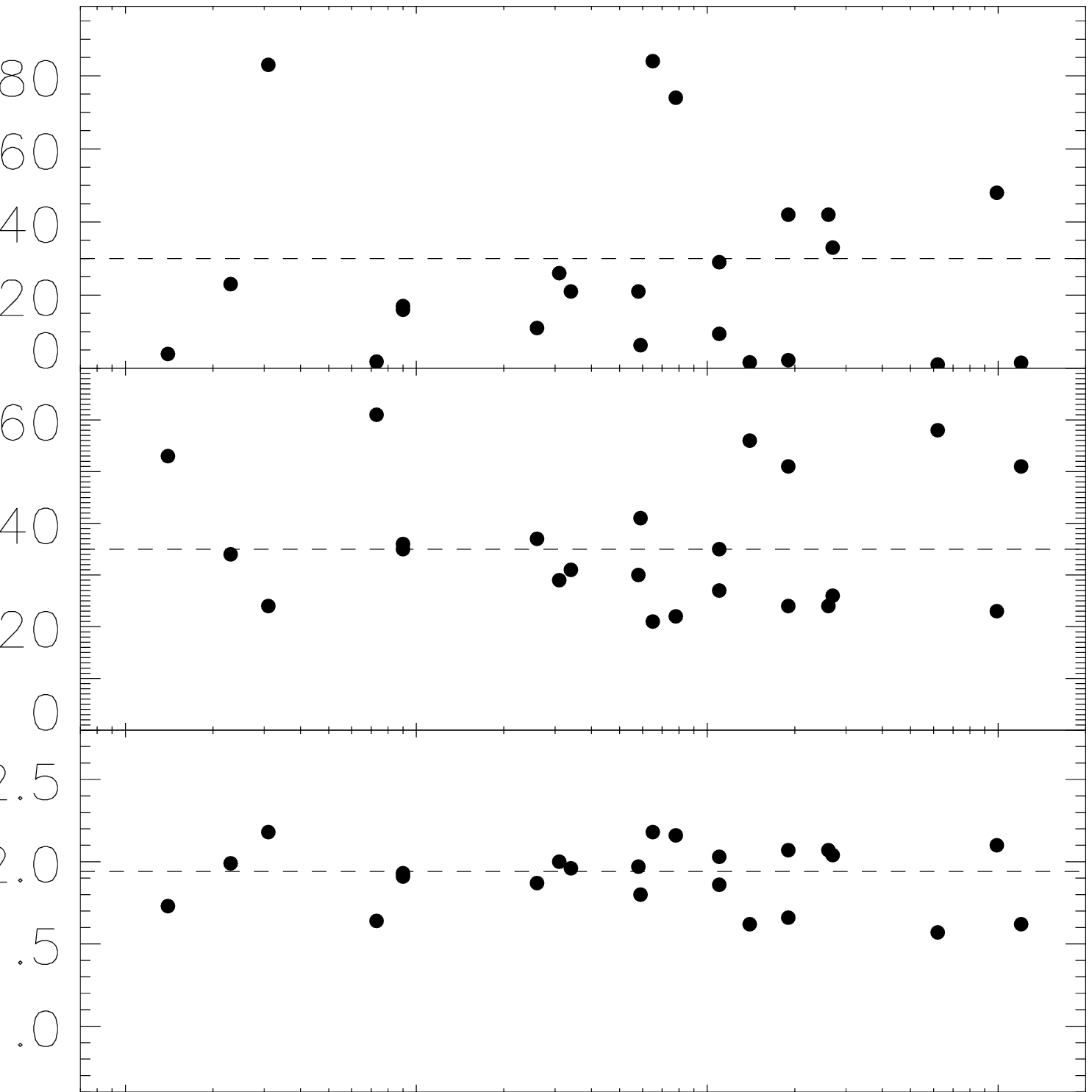}
\vspace{1cm}
\caption{For 23 dSph galaxies functions $P,\,S/M_6^{0.87},$
and $\eta_6=z_{f}M_6^{0.1}$ are plotted vs. the masses $M_6$.
Fits (\ref{pc3})\,\&\, (\ref{zcr4}) are plotted by dashed
lines
}
\label{23dsph}
\end{figure}

Thus for these galaxies we have for the central pressure, $P_c$,
baryonic density, $n_b$, and entropy, $S_b$
\[
\langle n_b\rangle\approx 28(1\pm 0.78)M_6^{-0.5}cm^{-3},\quad
\langle\eta\rangle\approx 3.2(1\pm 0.1)\,,
\]
\be
\langle P_c\rangle\approx 28(1\pm 0.9)eV/cm^3\,,
\label{pc3}
\ee
\[
\langle S_b\rangle = 36(1\pm 0.35)M_6^{0.87}eV\cdot cm^2\,.
\]
It is important that in spite of large scatter of measured
characteristics for these galaxies the pressure $P_c$ (\ref{pc3})
is quite similar to the pressure (\ref{p18},\,\ref{pc0},\,
\ref{pc1}\,\&\,\ref{pc2}) found above for clusters of
galaxies. As was discussed in Demia\'nski \& Doroshkevich
(2014) these results agree well with expectations of
simulations (Klypin et al., 2011) and reflect some important
intrinsic properties of violent relaxation and formation of
DM dominated virialized objects.

For the redshift of formation of the dSph galaxies we get
(Demia\'nski \& Doroshkevich 2014)
\be
\langle 1+z_{f}\rangle\approx 3.3(1\pm 0.12)M_{13}^{-0.1}\,,
\label{zcr4}
\ee
\[
\langle B^{-1}(z_{f})\rangle\approx 2.4(1\pm 0.12)M_{13}^{-0.1}\,.
\]

\subsection{Direct estimates of mass of the DM particles}

For  clusters of galaxies the thermal velocities in central
regions are $v_c\sim 100 - 1000 km/s$ and they are clearly
generated in the course of violent relaxation of the compressed
matter. In contrast, the observed velocity dispersion of the
dSph galaxies is not so large, $\sigma_{obs}\leq 10km/s$, what
allows to obtain direct rough estimates of the velocity,
free--streaming scale and mass of DM particles accumulated by
these galaxies. One example of such estimates can be found
in Boyarsky et al. (2009d) where for the mass of WDM particle
four values were obtained in the range
\be
m_{wdm}\geq 0.4 - 2.8 keV\,.
\label{mwdm-bb}
\ee
Here we can get similar estimates using properties of low
mass dSph galaxies with minimal velocity dispersions
$\sigma_{obs}\sim 3 - 4km/s$.

Assuming that formation of these galaxies is accompanied
by the adiabatic compression of weakly perturbed DM we can
estimate the random velocities of the same population of DM
particles before compression, $\sigma_{homo}$. For the four low
mass dSph galaxies this velocity dispersion is
\be
\sigma_{homo}(z=0)=\sigma_{obs}\left(\frac{\langle\rho_m(z=0)
\rangle}{\rho_c}\right)^{1/3}\sim 0.01km/s\,,
\label{vhom}
\ee
and their comoving radius is
\be
R_{homo}=(3M/4\pi\langle\rho_m(z=0)\rangle)^{1/3}\approx 11
kpc\,.
\label{velo}
\ee

The mass of these DM particles can be estimated from the
temperature at the redshift when these particles become
non relativistic (Doroshkevich et al. 1980):
\be
m_vc^2\sim 3.5kT_\gamma\frac{c}{\sigma_{homo}}\sim 22keV\,.
\label{mv}
\ee
According to the estimates of Bardeen et al. (1986) for such
particles the free--streaming scale is
\be
R_f\sim (50 - 100)kpc(1 keV/m_v)\sim (2 - 5)kpc\,,
\label{Rf}
\ee
what is even less than (\ref{velo}).
The low precision of measurements of both $\sigma_{obs}$ and
$\rho_c$ as well as a possible growth of $\sigma_{obs}$ owing
to the violent relaxation makes the  estimates (\ref{vhom}) and
(\ref{mv}) very rough.  Non the less, they demonstrate that
probably among the population of cosmological DM particles
there is a subpopulation with the mass (\ref{mv}) and the
damping scale (\ref{Rf}).

\subsection{Observed properties of the THING and LSB galaxies}
The number of observed DM dominated galaxies is very 
small and therefore we will use observations of all objects
for which the influence of DM component is significant.

Here we consider 14 THING
galaxies (de Blok et al. 2008) for which the contribution of
DM component seems to be important. The virial mass of
these galaxies can be roughly found from published rotation
curves while estimates of their central density are given in
de Blok et al. (2008). Similar estimates of mass and density
of these galaxies are presented also in Chemin et al. (2011).
However the central temperature of these galaxies is not
known and further analysis is based on estimates (\ref{dnfw},
\ref{pcc}). Using relation (\ref{dnfw}) we can estimate the
parameter $\eta$ and redshift $z_f$ for these 14 galaxies  with
the virial masses $ 5\cdot 10^9\leq M_{vir}/M_\odot\leq 7\cdot
10^{11}\,:$
\[
\langle\rho_c\rangle\approx 3.2\cdot 10^{-2}(
1\pm 0.8)M_\odot/pc^3,\quad
\langle \eta\rangle\approx 3.0(1\pm 0.1)\,,
\]
\be
\langle 1+z_f\rangle\approx 5.0(1\pm 0.2),\quad
\langle B^{-1}(M_{vir})\rangle\approx 3.7(1\pm 0.2)\,.
\label{thing}
\ee
The central pressure and entropy for this sample
can be found with (\ref{pcc}) with large scatter
\be
\langle P_c\rangle\approx 37(1\pm 0.9) eV/cm^3\,,
\label{pthings}
\ee
\[
\langle S_c\rangle\approx 27 cm^2eV(1\pm 0.9)\,,
\]
which can be partly related to the stronger
influence of the cooling process of baryonic component.

For the sample of LSB galaxies the masses and central densities
are discussed in Kuzio de Naray et al. (2008). As before for
these galaxies the central temperature was not measured and
our analysis is based on estimates (\ref{dnfw}, \ref{pcc}). For
10 LSB galaxies with $10^9\leq M_{vir}/M_\odot\leq 2\cdot
10^{11}$ we get
\[
\langle\rho_c\rangle\approx 2.7\cdot 10^{-2}M_\odot/pc^3
(1\pm 0.75),\quad
\langle \eta\rangle\approx 3.1(1\pm 0.1)\,,
\]
\be
\langle 1+z_f\rangle\approx 5.9(1\pm 0.1),\quad
\langle B^{-1}(M_{vir})\rangle\approx 4.4(1\pm 0.1)\,.
\label{lsb}
\ee
Here the parameter $\eta$ is determined by (\ref{eta})
and it coincides with the expected one. For this sample the
central pressure and entropy can also be found with
(\ref{pcc}) with large scatter
\be
\langle P_c\rangle\approx 14(1\pm 0.8) eV/cm^3\,,
\label{plsb}
\ee
\[
\langle S_c\rangle\approx 19cm^2eV(1\pm 0.7)\,,
\]
which also can be partly related to the stronger
random influence of the cooling process of  baryonic component.

It is necessary to bear in mind that for these objects the
virial mass is under estimated while the central density and
$z_f$ are over estimated owing to the
possible excess of baryons. Non the less the obtained value
of $\langle \eta\rangle$ (\ref{thing}) and (\ref{lsb}) shows
that the final results are sufficiently reasonable.

These results for both THING and LSB galaxies are plotted
in Fig. \ref{pds}\,\&\,\ref{sig6}.

\section{Central pressure and entropy in DM dominated
virialized objects}

\begin{figure}
\centering
\epsfxsize=6.5cm
\epsfbox{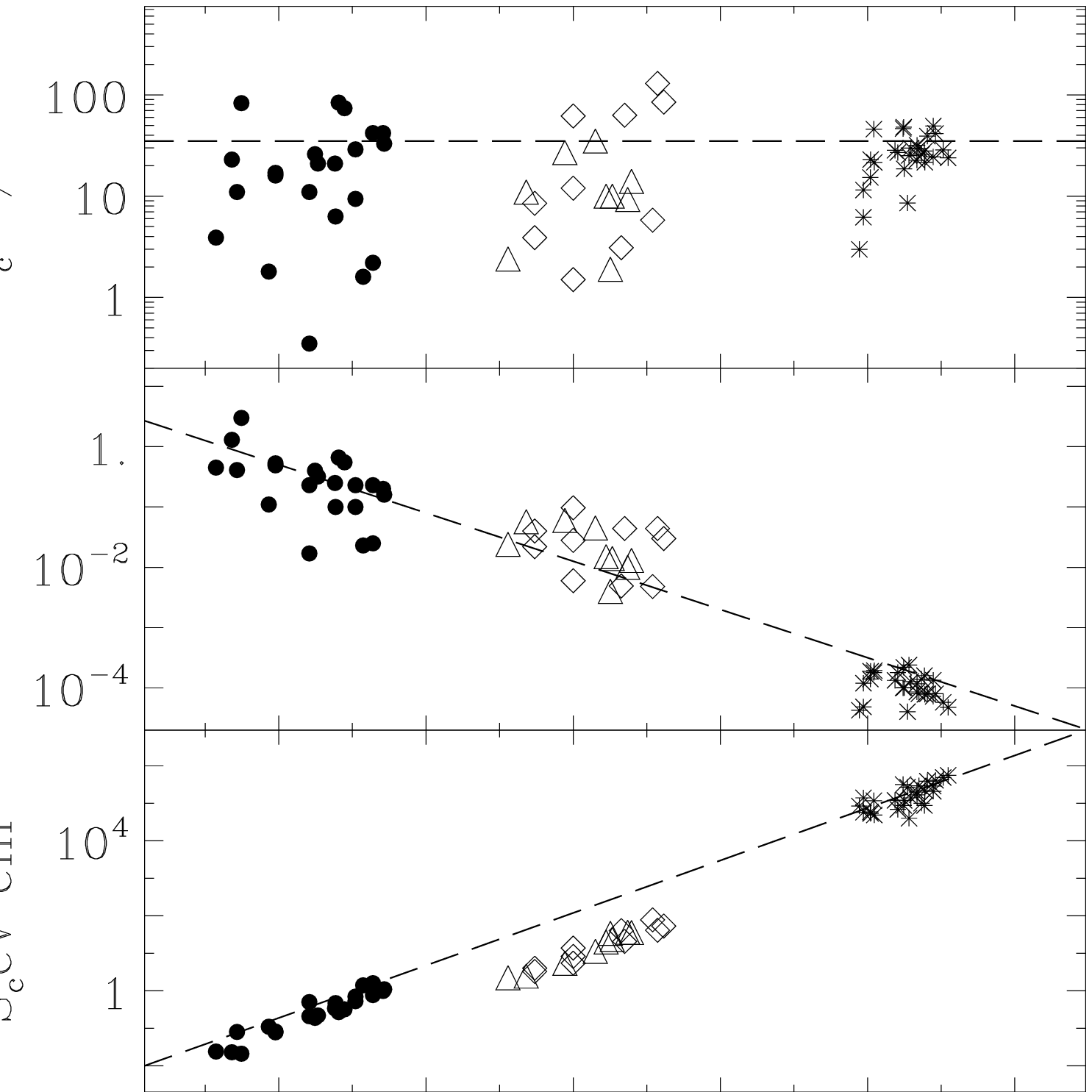}
\vspace{1cm}
\caption{The central pressure, density and entropy of 
virialized DM dominated objects are plotted vs. mass 
for dSph (points), THING (triangle) and LSB (rhombus) 
galaxies and for 9 SPT and 18 nearby (\ref{p18},\,
\ref{s18}) clusters of galaxies (stars) with x-ray 
masses and hot baryonic component. Fits (\ref{fit-pds})
are plotted by dashed lines.
}
\label{pds}
\end{figure}

Our analysis revealed some
unexpected peculiarities in the internal structure of DM
dominated virialized objects. First of them is  a very
weak dependence of the central pressure of such halos
(\ref{p18},\,\ref{pc0},\,\ref{pc1},\,\ref{pc2},\,\ref{boxsz},\,
\ref{pc3},\,\ref{pthings},\,\&\,\ref{plsb}) on the virial mass,
redshift of formation and other characteristics of these
objects. In contrast, the central entropy of clusters of
galaxies (\ref{s18},\,\ref{hot},\,\ref{cold}) significantly
exceeds the entropy of DM dominated objects of galactic
scale (\ref{pc3},\,\ref{pthings},\,\ref{plsb}).  For the
dSph, THING and LSB galaxies and for 9 SPT (\ref{pc1})
and 18 nearby (\ref{p18}, \ref{s18}) clusters of galaxies
 the central observed pressure, $P_c$, density, $\rho_c$,
 and entropy, $S_c$ are plotted in Fig. \ref{pds} and
 fitted by (\ref{fit-pds}) :
\be
P_c\approx 36eV/cm^3,\quad \rho_c\approx 2.\cdot 10^{-3}
M_{12}^{-0.4}M_\odot/pc^3\,,
\label{fit-pds}
\ee
\[
 S_c\approx 3 M_{12}^{0.7}keV cm^2,\quad M_{12}=
 M_{vir}/10^{12} M_\odot\,.
\]

The strong correlation of $n_c \& T_c$ together with weak
correlation of $P_c\,\&\,S_c$ discussed in Sec. 4.3.2
confirm the objective character of these inferences

Of course these effects have only statistical significance
what implies natural scatter of measured central pressures
and entropy and the parameter $\eta$ (\ref{eta}). However 
both the limited precision of observations and the impact 
of cooling process of baryonic component significantly 
increase the random scatter of real measurements.

In order to explain so different behavior of central pressure and
entropy it is necessary to remind that if the central pressure
is determined mainly by the dynamical equilibrium of compressed
DM  dominated component then the central entropy includes two
components, namely, the initial entropy of compressed matter and
entropy generated in the course of compression and relaxation.
The relative contribution of these components depends upon the
halo mass and redshift of formation and our results indicate that
the contribution of the first component progressively increases
together with the virial mass of formed halos. This inference
agrees with the well known regular growth of entropy with
radius within virialized objects, $S\propto M_{vir} r$,  what
indicates the progressive generation of entropy in the course
of matter relaxation. It is quite interesting to trace this
behavior in more details in both observed and simulated clusters.

Thus first galaxies such as dSph ones are formed from cold DM
particles and baryonic component  with very low entropy (see,
e.g. Demia\'nski \& Doroshkevich 2014). This implies that for
these objects the central entropy of both DM and baryonic
components (\ref{pc3}) is mainly generated in the course of
objects formation. In contrast, for later formed massive clusters
of galaxies the contribution of initial entropy progressively
increases and becomes more and more essential.

The initial entropy of baryonic component can be partly
related to the progressive heating of intergalactic gas by
ionizing UV background. For redshifts $z\leq 3$ when
strong photo ionization of HeI and HeII is  caused by the
hard UV radiation of quasars the entropy for slightly
perturbed baryons and 3D Hubble expansion can be
estimated as
\[
T_{b}\sim 0.7eV(1+z)^{6/7},\quad S_{b}\sim
18(1+z)^{-8/7}cm^2keV\,.
\]
This entropy strongly exceeds the entropy of the THING and 
LSB galaxies and is more similar to the observed entropy of
low mass clusters of galaxies. The Jeans mass of such 
baryonic component increases up to
\[
M_J\approx 10^6M_\odot\left (\frac{T}{10^4K}\right)^{3/2}
\left(\frac{1cm^{-3}}{\langle n_{bar}\rangle}\right)^{1/2}
\sim 10^{10}M_\odot (1+z)^{0.2}\,,
\]
and the formation of less massive objects is sharply decelerated.

At redshifts $z_f\geq 3$ the quasars contribution to 
ionizing UV radiation is small and the generated entropy of
baryons depends upon the shape of more soft spectrum of 
the UV background. Thus the observed entropy of THING
(\ref{pthings}) and LSB (\ref{plsb}) galaxies is similar 
to the entropy of dSph galaxies (\ref{pc3}) what shows that
for them the contribution of initial entropy is small. 
This means that probably at redshifts $z_f\geq 3$ the ionization
of intergalactic gas is not accompanied by its essential heating.

On the other hand extraordinary efforts are required in order
to increase the much more conservative entropy of DM
component. As was noted above, perhaps, the progressive
growth of small scale perturbations and their following
randomization is the most promising way to increase the
large scale entropy of both DM and baryonic components
and to explain the observed high entropy of clusters of
galaxies. The problem requires further analysis in simulations.

The high stability of the central pressure for DM dominated
objects was already noted in our previous paper (Demia\'nski
\& Doroshkevich 2014) and here it is confirmed with wider
observational base. It can be related to the combined influence
of the violent relaxation and of the regular shape of the initial
power spectrum of density perturbations and so, also velocity
perturbations.

It is important that for the CDM model  simulations show
similarity of the dimensionless characteristics of DM halos
such as the density and pressure profiles.  In particular
the density profiles are found to be close to NFW or Burkert
(1995) ones with moderate variations of concentration $C\sim
3 - 5$. On the other hand the regular character of the CDM
(Bardeen et al. 1986) and WDM (\ref{viel}) initial power
spectra links together the virial mass of DM objects with their
redshifts of formation and mean densities as this is demonstrated
by Eqs. (\ref{eta}, \ref{dnfw}, \ref{pcc}) and is discussed in
the next Section. So the impact of the growth with time of the
virial mass of formed objects is compensated approximately by
corresponding drop of their mean density (see, e.g.,
(\ref{rho_mns},\,\ref{dnfw})).

\section{MDM cosmological models}

Simulations show that characteristics of the virialized DM
halos are much more stable than the characteristics of baryonic
component and after formation at $z=z_{f}$ of virialized DM
halos with $\langle\rho_{vir}\rangle\approx 18\pi^2\langle\rho(
z_{f})\rangle$ slow matter accretion only moderately changes
their characteristics (see, e.g., Diemer et al. 2013). Because
of this, we can observe earlier formed high density galaxies
with moderate masses even within later formed more massive but
less dense clusters of galaxies, filaments and other elements
of the Large Scale Structure. This means that using the model
presented in Sec. 3 for description of the observed dSph,
THING and LSB galaxies and clusters of galaxies dominated by
DM component we can find one--to--one correspondence between
their observed parameters and the so called redshift of object
formation, $z_{f}$. Of course according to the Press --
Schechter approach (Press, Schechter, 1974; Peebles 1974;
Bond et al. 1991) these redshifts characterize the power
spectrum of the density perturbation rather than the real
period of the object formation.

\begin{figure}
\centering
\epsfxsize=6.5cm
\epsfbox{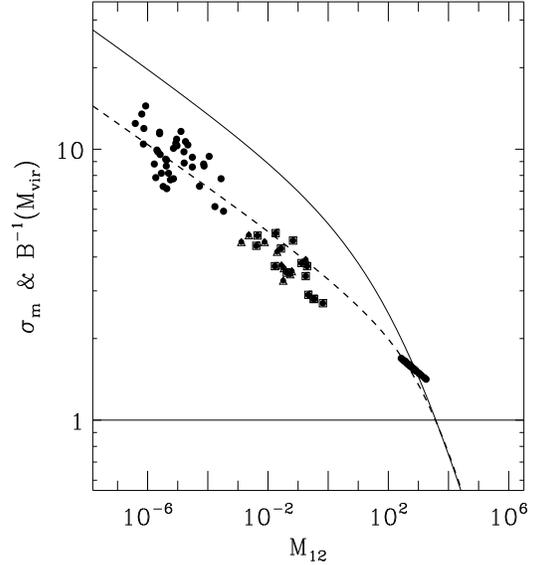}
\vspace{1cm}
\caption{Correlation function of the matter density $\sigma_m$
for the CDM power spectrum  (\ref{fcdm})  and the combined
spectrum (\ref{best}) are plotted by solid and dashed
lines. Function $B^{-1}(M_{vir})$ is plotted for the 41 dSph
galaxies (left group of points) and for 31 clusters of
galaxies from the SPT observed sample (right group of
points). For 14 THINGS galaxies function $B^{-1}(M_{vir})$
is plotted by squares and for 10 LSB galaxies - by triangles.
}
\label{sig6}
\end{figure}

Following the Press -- Schechter ideas (\ref{bsig}) we
compare the function $\sigma_m(M)$ for various DM models with
the observed function $B^{-1}(M_{vir})$ obtained in the
previous Sections in a wide range of masses and redshifts. Such
comparison allows us to quantify the influence of the DM
particles on the rate of DM halos formation. Thus in Fig.
\ref{sig6} the function $B^{-1}(M_{vir})$ is plotted for
the dSph, THING and LSB galaxies and for the sample of 31
clusters of galaxies (\ref{zcr3}) together with the function
$\sigma_m(M_{vir})$ calculated for the standard CDM power
spectrum, and for the MDM power spectrum (\ref{best}) with
$\sigma_m$ obtained according to (\ref{fmdm}) for low massive
WDM particles
\be
p(k)=0.27p_{cdm}+0.73p_{wdm}(m_{w}),\quad m_w\approx 30 eV\,,
\label{best}
\ee
and corresponding damping scale $M_{dmp}\sim 10^{14}
M_\odot$. The functions $\sigma_m(M_{vir})$ are normalized
using the cluster points (\ref{zcr3}).

Of course the observational base used in this discussion is
very limited and it should be extended by adding observations
of new objects with masses $M\simeq 10^{10} - 10^{12} M_\odot$,
what may be crucial for determination of the real composition
of dark matter. Unfortunately more or less appropriate
estimates of the redshift $z_{f}$ can be obtained mainly for
DM dominated objects or for objects with clearly discriminated
impact of DM and baryonic components. Here we use
results obtained for 14 THINGS galaxies (de Blok et al. 2008)
and 10 LSB galaxies (Kuzio de Naray et al. 2008). We hope
that the list of possible appropriate candidates can be
extended.

Next important problem is the reliability of the approach
used in our analysis and obtained results. It depends upon 
the representativity of the observational data and is low
because of the very limited available data and their
significant scatter. Progress achieved during last years
allows to begin discussion of MDM models but the low
reliability of available observations is causing only 
qualitative character of our discussion.  Indeed, the 
problem of estimates of the mass, density and other 
parameters of the observed objects is quite complex, 
methods used for such estimates are very rough and model 
dependent while their reliability is limited. Moreover 
the influence of baryonic component increases scatter 
of the measured parameters and makes it difficult to 
estimate the real precision of our results. Let us hope 
that because of the importance of this problem such 
observations will be extended and their precision improved.

On the other hand the simulated data are focused on 
clusters of galaxies, what is caused by the finite 
resolution of simulations. This limitation also does
not increase reliability of our inferences. Non the less let
us note that the analysis discussed in Sec. 4 is based on
results of very large and high quality simulations of the
standard $\Lambda$CDM model (Klypin
et al. 2011) which embraces relatively large accessible
range of object masses and redshifts of formation.

As is seen from Figure \ref{sig6} the standard CDM model
cannot describe the system of observed points and, so,
should be rejected. Quality of the WDM model with
$m_w\approx 3 keV$ is also limited and simulations
(Shultz et al. 2014) show that in this model the objects
formation at high redshifts is oversuppressed.
Better description of observations is presented in Fig.
\ref{sig6} by the two component DM model with parameters
(\ref{best}).

The important result of our analysis is the demonstration of
limited applicability of cosmological models with only one
component power spectrum and great promises of models with
more complex structure of the power spectrum. It is also
important that the basic element of such complex power
spectrum is the large contribution ($\sim 70\%$) of the
low mass WDM spectrum with $m_w\sim 30 eV$. It is noteworthy
that after 30 years of absolute domination of the CDM model
we return to more complex versions of the HDM models.

In turn such models imply existence of at least two
damping scales one of which corresponds to the mass of
clusters of galaxies, $M_{dmp}\sim 10^{13} M_\odot$.
However such partial damping of the power spectrum leads
only to a decrease of the rate of formation of less
massive objects relatively to the standard CDM--like
power spectrum.

Of course all these inferences are very preliminary. Thus,
here we use the WDM power spectra with the transfer
function (\ref{viel})\, (Viel et al. 2005; Polisensky \&
Ricotti 2011; Marcovi$\breve{c}$ \& Viel  2013). More refined
description of the MDM power spectrum and specially
further progress in observations of DM dominated objects
will change the best model parameters and estimates of
masses and composition of the MDM model. Non the less
even today replacement of some fraction of CDM particles
by heavy WDM particles with
$m_w\sim 10 keV$ can be considered. Spectrum of such
particles identified now (according to majority preference)
as the sterile neutrinos can be included in (\ref{best}) as
a third component without noticeable changes of Fig.
\ref{sig6}. Indeed, the available sample of observed
DM dominated objects does not yet allow us to make any
far--reaching conclusions about the actual properties of
massive DM particles. However, the noticeable contribution
of low mass WDM spectrum in (\ref{best}) is crucial for the
considered models

At the same time it is very important that the observed
characteristics of objects are determined mainly by the
linear combination of power spectra and so by the damping
scales which in turn depend both upon particle masses and
velocities. This means that the construction of the
adequate complex cosmological model should include discussion
of full DM evolution beginning from the period of inflation
with estimates of the actual damping scales and transfer
functions for all components allowing also for linear
evolution of perturbations at $z\leq z_{eq}$. Simple
example of such evolution is discussed by Boyarsky et al.
(2009b) and in Sec. 2.3\,.

Thus, allowing for the decay of the separate components of
power spectra according to (\ref{decay}) we can estimate the
matter fractions of CDM and WDM components, $f_{cdm}\,\&\,
f_{wdm}$, for the model (\ref{best}) with $g_{cdm}=0.27$
\be
f_{cdm}\approx 0.82,\quad f_{wdm}\approx 0.18\,.
\label{fm2}
\ee
Unexpectedly in spite of the
relatively small value of $g_{cdm}$ the fraction of the CDM
particles, $f_{cdm}$, remains significant as well as their
influence upon the evolution of small scale objects. However
the widely discussed controversial characteristics of DM
objects such as the core--cusp problem, or number of low
mass satellites depend upon the dissipative scale and the
power spectrum rather than directly upon the mass or fraction
of the CDM component. This supports the hope that these
simpler problems also will be successfully resolved in the
framework of the discussed MDM cosmological models.

These discussed complex intercorrelations produce additional
problems for numerical simulations which now practically
cannot simultaneously provide suitable size of computational
box and required high mass resolution. This means that direct
simulations with realistic complex power spectrum encounter
many problems and require to use model simulations with
subsequent rescaling procedures what makes difficult further
comparison with observations. As usual the additional
problem is the accurate description of possible impact of
baryonic component.

So, more accurate simulations with various compositions of
dark matter are required before we will have reliable
inferences about properties of the DM component.
These doubts are supported by moderate results of the first
published simulations of the WDM cosmological models
(Maccio 2012, 2013;  Angulo, Hahn, Abel, 2013; Schneider,
Smith \& Reed, 2013; Wang et al. 2013; Libeskind et al. 2013;
Marcovi$\breve{c}$ \& Viel 2013; Schultz et al. 2014;
Schneider et al. 2014; Dutton et al. 2014).

\section{Conclusions}

In this paper we discuss two important problems of modern
cosmology:
\begin{enumerate}
\item{}the composition of the dark matter,
\item{} the internal structure of virialized DM dominated
objects  -- the high stability of their central pressure and
dependence of the central entropy upon mass and period
of halos formation.
\end{enumerate}

Summing up it is necessary to note that the proposed approach 
allows us to consider and to compare properties of DM
dominated objects in an unprecedentedly wide range of 
masses $10^5\leq M_{vir}/M_\odot\leq 10^{15}$. This comparison
unexpectedly favors MDM models for which the domination 
of massive DM component is accompanied by a significant
contribution of low mass DM particles. In turn in these 
models the power spectra of density perturbations at galactic
scale significantly differ from the standard CDM--like ones.

Here we consider as a quite promising the MDM model
for which the power spectrum is composed of fraction
$g_{cdm}\sim 0.3$ of the CDM spectrum and fraction
$g_{wdm}\sim 0.7$ of the WDM spectrum with low mass
thermalized WDM particles and transfer function (\ref{viel}).
The rough estimates of the mass fraction of these components
are given by (\ref{fm2}). Further progress can be achieved
with more complex models with more realistic transfer functions
instead of (\ref{viel},\,\ref{fmdm}) and/or with replacement
of some fraction of CDM particles by
heavy WDM particles with mass $m_w\geq 3 keV$ (such
as the sterile neutrino).

We want to emphasize that unexpectedly in the spectrum 
(\ref{best}) of this more promising MDM model the spectrum 
of low mass WDM particles with relatively large damping scale 
dominates. This can be considered as reincarnation in a new
version of the earlier rejected HDM model. It is important 
that in the MDM model the impact of such low mass WDM 
particles decelerates the growth of perturbations and the 
rate of the objects formation for all objects with masses
less than the low mass clusters of galaxies or massive 
galaxies, $M_{vir}\leq 10^{12} M_\odot$. However the
contribution of CDM--like spectrum provides successful 
formation of low mass halos and other structure elements.

Further development of this approach could result in 
further complication of power spectra and in particular 
in introduction of an excess of power localized at small 
scale. Such modifications allow to essentially extend
possibilities of the MDM models and even can allow to 
link  the possible unexpected observed properties of some 
set of objects of galactic scales with peculiarities of 
the power spectrum. In some respect such an excess of 
power reminds the isocurvature models (see, e.g. Savelainen 
et al. 2013) with similar predictions and problems. However 
all such problematic multi parametric proposals should be 
considered in the context of general cosmological and 
inflationary models.

During the last decade the most reliable and interesting 
information about the power spectrum comes from observations
of the perturbations at the period of recombination which 
are seen as the CMB fluctuations. Such observations are
performed both with satellites (Komatsu et al. 2011; Larson 
et al. 2011; Ade et al. 2013) and the SPT and other ground
telescopes (see, e.g., Saro et al. 2013). But they relate 
to large scales $L\geq 10Mpc$ only. Limited information 
about the power spectrum at smaller scales can be obtained 
from observations of absorption spectra of distant quasars. 
But properties of such absorbers are very sensitive to 
spatial variations of poorly known ionizing UV background, 
what makes difficult interpretation of these observations 
and strongly limits their reliability.

The approach used in this paper for discussion of composition
and properties of the DM particles is very indirect.
We  consider effects of strongly nonlinear multistep evolution
of perturbations resulting in observed properties of the
relaxed objects. The nonlinear evolution already leads to a
strong loss of information about the primeval perturbations
and composition of the DM component. These losses are
further enhanced by the masking effects of dissipative
evolution of the baryonic matter. To seek out the missing
impact of the DM composition and primeval perturbations we
have to compare observational data with numerical
simulations majority of which are now focused on studying
evolution of the standard $\Lambda$CDM models.

We used the general theory of gravitational instability for
the DM objects to justify the expression (\ref{bsig}) and
Figure \ref{sig6} and to quantify correlations between
the virial mass of objects, $M_{vir}$, their redshift of
formation, $z_{f}$, and the shape of the primordial
power spectrum of perturbations (\ref{fmdm}) in a wide
range of masses. In this respect our approach seems to
be more helpful and informative than the earlier
mentioned contradictions between the observed and
simulated characteristics of the DM halos. Non the less
it provides us with only preliminary qualitative inferences
about the nature of DM particles and cosmological models.

It can be expected that another manifestation of the same
interactions is the well known similarity of the internal
structure of DM dominated virialized objects and, in
particular, the high stability of the central pressure
of these objects discussed in Secs. 4 \& 6.

As is seen from Figs. \ref{crw} and \ref{sig6} deviations 
between the CDM and MDM power spectra appear at scales
$M_{vir}/M_\odot\sim 10^{12} - 10^{13}$. This means that 
properties of massive galaxies and clusters of galaxies such
as their central pressure, entropy and specially the rate 
of formation are more sensitive to the compositio
n of the
dark matter and for impact of low mass particles. These 
possibilities were discussed in Sec. 6. Unfortunately 
published -- and discussed in Secs. 4 and 5 -- observations 
of such objects are limited and their interpretation 
is unreliable what do not allow us to reveal the expected 
effects. Additional problem is the impact of cooling of 
the baryonic component which masks the possible impact 
of the power spectrum and DM composition.

We believe that more detailed conclusions will be made
on the basis of special simulations and after accumulation
of more representative observational sample of high precision
measurements of properties of DM dominated objects.

\subsection{Acknowledgments}
This paper was supported in part by  the
grant of president RF for support of scientific schools
NS4235.2014.2. We thank S.Pilipenko, B.Komberg, A.Saburova
and R.Rufini for useful comments. We wish to thank the 
anonymous referee for many valuable comments and suggestions.


\begin{thebibliography}{}

\bibitem[2013]{Ade 2013}
Ade, P., et al. 2013, arXiv:1303.5076
\bibitem[2013]{Agnese 2013}
Agnese, R., et al. 2013, arXiv:1305.2405

\bibitem[2012]{ander}
Anderhalden, D., Diemand, J., Bertone, G., Maccio, A.,
Shneider A., 2012, JCAP, 10, 047

\bibitem[2013]{Anglo}
Angloher, G., et al., 2013, Eur.Phys.J., C72, 1971

\bibitem[2005]{Arnaud 2005}
Arnaud, M., Pointecouteau, E., Pratt, G., 2005, A\&A, 441, 893

\bibitem[2010]{Arnaud 2010}
Arnaud, M., Pratt, G., Piffaretti, R., B\"oringer, H.,
Croston, J., Pointecouteau, E., 2010, A\&A, 517, 92;
arXive:0910.1234

\bibitem[2012]{angulo 2012}
Angulo, R., Springel, V., White, S.D.M., Jenkins, A., Baugh, C.,
Frenk, C., 2012, MNRAS, 426, 2046

\bibitem[2013]{angulo 2013}
Angulo, R., Hahn, O., Abel, T., 2013, arXiv:1304.2406

\bibitem[1986]{BBKS}
Bardeen~J.M., Bond~J.R., Kaiser~N., \& Szalay~A., 1986,
ApJ, 304, 15

\bibitem[2003]{Benn}
Bennet C., et al., 2003, ApJS,, 148, 1

\bibitem[2013]{Bern}
Bernabei, R., et al., 2008, Eur.Phys.J., C56, 333

\bibitem[2013]{Bern}
Bernabei, R., et al., 2010, Eur.Phys.J., C67, 39

\bibitem[1984]{blum 1984}
Blumenthale, G., Faber, S., Primck, J., Rees, M., 1984,
Nature, 311, 517.

\bibitem[2013]{Bhatta-13}
Bhattacharya, S., Habib, S., Heitmann, K., Vikhlinin, A., 2013,
ApJ, 766, 32 

\bibitem[1980]{Bisn}
Bisnovaty-Kogan, G., \& Novikov, I., 1980, Astr.Zh., 57, 899

\bibitem[1996]{bode 2001}
Bode,P., Ostriker, J., Turok, N., 2001, ApJ, 556, 93

\bibitem[1980]{bond 1980}
Bond, R., Efstathiou, G., Silk, J., 1980, Phys.Rev.Lett., 45, 1980

\bibitem[1983]{bond 1983}
Bond, R., Szalay, A., 1983, ApJ, 274, 443

\bibitem[1984]{bond 1984}
Bond R. \& Szalay A., 1984, ApJ, 274, 443;

\bibitem[1991]{bond 1991}
Bond, R., Cole, S., Efstathiou, G., Kaiser, N., 1991, ApJ, 
379, 440

\bibitem[2014]{Borde-14}
Borde, A., Palanque-Delabrouille, N., Rossi, G., Viel, M.,
Bolton, J., Yeche, C., Le Goff, J.-M., Rich, J, 2014,
arXiv:1401.6472

\bibitem[2009]{bovill 2009}
Bovill, M, Ricotti, M., 2009, ApJ, 693, 1859

\bibitem[2009]{boy 2009}
Boyarsky, A., Lesgourgues, J., Ruchayskiy, O., Viel, M., 2009a,
JCAP, 05, 012

\bibitem[2009]{bo 2009}
Boyarsky, A., Lesgourgues, J., Ruchayskiy, O., Viel, M., 2009b,
PhRvL, 102t1304

\bibitem[2009]{boya 2009}
Boyarsky, A., Ruchayskiy, O., Shaposhnikov, M., 2009c,
ARNPS, 59, 191

\bibitem[2013]{boya 2013}
Boyarsky, A.,  Ruchayskiy, O., Iakubovskiy D.,2009d,
JCAP, 0903, 005

\bibitem[2013]{boya 2013}
Boyarsky, A., Iakubovskiy D., Ruchayskiy, O., 2013,
arXiv:1306.4954

\bibitem[2014]{boya 2014}
Boyarsky, A., Ruchayskiy, O., Iakubovskiy D., Franse, J.,
2014, arXiv:1402.4119

\bibitem[2012]{boyl 2012}
Boylan-Kolchin, M., Bullok, J., Kaplinghat, M., 2012, MNRAS,
422, 1203

\bibitem[2007]{Branchesi}
Branchesi, M., Gioia, I., Fanti, C.,Fanti, R., 2007, A\&A,
472, 739

\bibitem[1998]{Bryan-n}
Bryan, G., Norman, M., 1998, ApJ, 495, 80

\bibitem[2014]{Bulbul}
Bulbul, E., Markevitch, M., Foster, A., Smith, R., Loewenstein,
M., Randall, S., 2014, arXiv:1402.2301

\bibitem[2001]{Bull}
Bullock, J., Kolatt, T., Sigad, Y., Somerville, R., Kravtsov, A.,
Klypin, A., Primack, J., Dekel, A., 2001, MNRAS, 321, 559

\bibitem[2012]{Bur}
Burenin, R., Vikhlinin, A., 2012, Astronomy Lett., 38, 1

\bibitem[1996]{Burk}
Burkert~A., 1995, ApJ, 447, L25

\bibitem[2014]{Carr 2014}
Carr, B., 2014, arXv:1402.1437

\bibitem[2011]{Chemin 2011}
Chemin, L., de Blok, W., Mamon, G., 2011, AJ, 142, 109

\bibitem[2013]{Choi 2013}
Choi, K.-Y., Kim, J., Roszkowski, L., 2013, 1307.3330  

\bibitem[2014]{Coll 2014}
Collins, M., et al., 2014, ApJ, 783, 7

\bibitem[2013]{costanzi}
Costanzi, M., Villaescusa-Navarro, F., Viel, M., Xia, J.-Q.,
Borgani, S., Castorina, E., Sefusatti, E., 2013, JCAP, 12, 012

\bibitem[2008]{croston}
Croston, J., et al., 2008, A\&A, 487, 431

\bibitem[2013]{cyr}
Cyr-Racine, F.-Y., Sigurdson, K., 2013, Phys.Rev.D, 87, 103515

\bibitem[2014]{day}
Dayland, T., Finkbeiner, P., Hooper, D., Linden, T., Portillo,
S., Rodd, N., Slatyer, T., 2014, arXiv:1402.6703

\bibitem[2008]{deblok}
de Blok, W., Walter, F., Brinks, E., Trachternach, C., Oh,
S.-H., Kennikutt, R., 2008, AJ, 136, 2648

\bibitem[1999]{demianski 1999}
Demia\'nski,M., Doroshkevich,A., 1999, ApJ, 512, 527

\bibitem[1999]{demianski 2004}
Demia\'nski,M., Doroshkevich,A., 2004, A\&A, 422, 423

\bibitem[2006]{dor 2006}
Demia\'nski, M., Doroshkevich A.,\& Turchaninov, V., 2006,
MNRAS, 372, 915

\bibitem[2011]{demianski 2011}
Demia\'nski,M., Doroshkevich,A., Pilipenko, S., Gottl\"ober, S.,
2011, MNRAS, 414, 1813.

\bibitem[2014]{dor 2014}
Demia\'nski, M., Doroshkevich A., 2014, MNRAS, 439, 179.

\bibitem[2007]{di 2007}
Diemand, J., Kuhlen, M., Madau, P., 2007, ApJ, 667, 859

\bibitem[2013]{diem 2013}
Diemer, B., More, S., Kravtsov, A., 2013, ApJ, 766, 25

\bibitem[2013]{dipak 2013}
Dipak, M., Cole, P., Viel, M., 2012, MNRAS, 427, 2359

\bibitem[1980]{dkk 1980}
Doroshkevich, A., Khlopov, M., Sunyaev, R., Zel'dovich, Ya.,
1980, SvAL,  6,  252

\bibitem[1981]{dkk 1981}
Doroshkevich, A., Khlopov, M., Sunyaev, R., Szalay, A.,
Zel'dovich, Ya., 1980, Proc. 10th Texas Symposium on Relativistic
Astrophysics, Ann.New York Acad. Sci., 375, 32.

\bibitem[1981]{dor 1981}
Doroshkevich A., Zel'dovich Ya., 1981, JETP, 80, 801

\bibitem[1984]{dk 1984}
Doroshkevich, A., Khlopov, M., 1984, MNRAS, 211, 277.

\bibitem[1986]{dkk 1986}
Doroshkevich, A., Khlopov, M., Kotok, E., 1986, SvA., 30, 251

\bibitem[1988]{dkk 1988}
Doroshkevich, A., Klypin, A., Khlopov, M., 1988, SvA., 32, 127

\bibitem[2003]{dor 2003}
 Doroshkevich A., Naselsky, I., Naselsky, P., Novikov, I., 2003,
ApJ., 596, 709

\bibitem[2013]{dor 2012}
 Doroshkevich A., Lukash, V., Mikheeva, E., 2012, PhyU., 55, 3

\bibitem[2013]{Dreves 2013}
Dreves, M., 2013, J.Mod.Phys. E,22,1330019

\bibitem[2014]{Dutton 2014}
Dutton, A., Maccio, A., 2014, arXiv:1402.7073

\bibitem[1998]{eisen}
Eisenstein, D., Hu, W., 1998, ApJ., 496, 605

\bibitem[2010]{feng 2010}
Feng, J., 2010, ARA\&A, 48, 495

\bibitem[2013]{hunter 2013}
Ferrer, F., Hunter, D., 2013, JCAP, 09, 005

\bibitem[1984]{Fillmore-Goldreich}
Fillmore~J.A., \& Goldreich~P., 1984, ApJ, 281, 1

\bibitem[2013]{FoH{e}x}
Fo\H{e}x et al., 2013, A\&A, in press, arXiv:1208.4026

\bibitem[2014]{Gait}
Gaitskell, R., 2014, ARNPS, 54, 315

\bibitem[1981]{grishchuk}
Grishchuk, L., Zeldovich, Ya., 1981, SvA., 25, 267

\bibitem[2012]{Gover}
Governato, F. et al., 2012, MNRAS, 422, 1231

\bibitem[1995]{Gurevich}
Gurevich, A., Zybin, K., 1995, Phys.Usp., 38, 687

\bibitem[2013]{hin 2013}
Hinshaw G., et al., 2013, ApJS, 208, 19

\bibitem[2013]{hori 2013}
Horiuchi, S., Humphrey P., Onorbe, J., Abazajian, K.,
Kaplinghat, M., Garrison-Kimmel, S., 2013, arXiv:1311.0282

\bibitem[2013]{knede}
Khedekar, S., Churazov, E., Kravtsov, A., Zhuravleva, E., 
Lau, E., Nagai, D., Sunyaev, R., 2013, MNRAS, 431, 954

\bibitem[2011]{klypin 2011}
Klypin, A. Trujillo-Gomez, S., Primack, J., 2011, ApJ, 740, 102;
arXiv:1002.3660

\bibitem[2014]{koll}
Kollmeier, J., et al., 2014, arXiv:1404.2933

\bibitem[2011]{komatsu-11}
Komatsu, E., et al. 2011, ApJS, 192, 18

\bibitem[2011]{komatsu-11}
Koposov, S., Yoo, J., Rix, H.W., Weinberg, D., Maccio, A.,
Escude, J., 2009, ApJ, 696, 2179

\bibitem[2012]{kravtsov 2012}
Kravtsov, A., Borgani, S., 2012, ARA\&A, 50, 353 

\bibitem[2009]{kus 2009}
Kusenko, A., Phys.Rep.,2009, 481, 1

\bibitem[2013]{abaz 2013}
Kusenko, A., Rosenberg, L., 2013, 1310.8642

\bibitem[2008]{kuzio 2008}
Kuzio de Naray, R., Mcgaugh, S., de Blok, W., 2008, ApJ, 
676, 920

\bibitem[2011]{laport-13}
Laporte, C.,. Walker, M., Penarrubia, J., 2013, MNRAS, 
433, 54L

\bibitem[2011]{larsen-11}
Larson, D.., et al. 2011, ApJS, 192, 16

\bibitem[2013]{Lib-14}
Libeskind, N., Di Cintio, A., Knebe, A., Yepes, G., Gottl\"ober,
S., Steinmetz, M., Hoffman, Y., Martinez-Vaquero, L., 2013,
arXiv:1305.5557

\bibitem[2011]{Lithwick-11}
Lithwick Y., Dalal N., 2011, ApJ, 734, 100L, arXiv:1010.3723;

\bibitem[2011]{Lloyd-davies}
Lloyd--Davies, E., 2011, MNRAS, 418, 14

\bibitem[2001]{Loeb-01}
Loeb, A., Barkana, R., 2001, ARA\&A, 39, 19

\bibitem[2013]{Ludlow-13}
Ludlow, A. et al., 2013, MNRAS, 432, 1103L 

\bibitem[2012]{maccio 2012}
Maccio, A., Paduroiu, S., Anderhalden, D., Schneider, A.,
Moor, B., 2012, MNRAS, 424, 1105

\bibitem[2013]{macci 2013}
Maccio A., Ruchayskiy O., Boyarsky A., Munos--Cuartas J., 
2013, MNRAS, 428, 882

\bibitem[2010]{Mantz}
Mantz, A., Allen, S., Rapetti, D., Ebeling, H., 2010, MNRAS,
406, 1759
\bibitem[2013]{marc 2013}
Marcovi$\breve{c}$, K., Viel, M., 2013, arXiv:1311.5223

\bibitem[2013]{McDonald}
McDonald, M., et al., 2013, ApJ, 774, 23

\bibitem[2014]{medved}
Medvedev, M., 2014, Phys.Rev.Let., 113, 071303

\bibitem[1999]{meik}
Meiksin, A., White, M., Peacock, J., 1999, MNRAS, 304, 851

\bibitem[2007]{dor 2007}
Mikheeva, E., Doroshkevich A., Lukash, V.,, 2007, NCimB,
122, 1393

\bibitem[2014]{mod 2014}
Modak, R., Majumdar, D., Rakashit, S., 2013, arXiv:1312.7488

\bibitem[2011]{Moughan-11}
Moughan, B., Giles, P., Randall, S., Jones, C., Formen, W.,
2011, arXiv:1108.1200

\bibitem[2007]{nagai-07}
Nagai, D., Kravtsov, A., Vikhlinin A., 2007, ApJ, 668, 1

\bibitem[1995]{NFW-95}
Navarro~J.F., Frenk~C.S., \& White~S.D.M., 1995, MNRAS, 275, 720

\bibitem[1996]{NFW-96}
Navarro~J.F., Frenk~C.S., \& White~S.D.M., 1996, ApJ, 462, 563

\bibitem[1997]{NFW-97}
Navarro~J.F., Frenk~C.S., \& White~S.D.M., 1997, ApJ, 490, 493

\bibitem[2010]{Nu-10}
Nulsen, P., Powell, S., Vikhlinin, A., 2010, ApJ, 722, 55

\bibitem[1990]{Peacock}
Peacock, j., Heavens, A., 1990, MNRAS, 243, 133

\bibitem[1967]{Peebles-67}
Peebles P.J.E., 1967, ApJ, 147, 859

\bibitem[1974]{Peebles-74}
Peebles P.J.E., 1974, ApJ, 189, L51

\bibitem[2000]{Peebles-20}
Peebles P.J.E., Seager, S., Hu, W., 2000, ApJ, 539, L1

\bibitem[2008]{Penar-08}
Penarrubia, J., Navarro, J., McConnachie, A., 2008, ApJ, 673, 
226

\bibitem[2010]{Penar-10}
Penarrubia, J., Pontzen, A., Walker, M., Koposov, S.,
2012, ApJ, 759L, 42

\bibitem[2012]{Penar-12}
Penarrubia, J., Benson, A., Walker, M., Gilmore, G., 
McConnachie, A., Mayer, L., 2010, MNRAS, 406, 1290

\bibitem[2012]{Pili}
Pilipenko, S., Doroshkevich, A., Lukash, V., Mikheeva, E.,
2012, MNRAS, 427L, 30

\bibitem[2006]{pointe_05}
Pointecouteau, E., Arnaud, M., Pratt, G., 2005, A\&A, 435, 1

\bibitem[2011]{poli}
Polisensky E. \& Ricotti M., 2011, PhRvD, 83.043506

\bibitem[2014]{pont}
Pontzen, A., \& Governato, F., 2014, Nature, 506, 171,

\bibitem[2010]{Prada 2012}
Prada, F., Klypin, A., Cuesta, A., Betancort-Rijo, J.,
Primack, J., 2012, MNRAS, 423, 3018   

\bibitem[2006]{pratt_06}
Pratt, G., Arnaud, M., Pointecouteau, E., 2006, A\&A, 446, 429

\bibitem[2009]{pratt_09}
Pratt, G., Croston, J., Arnaud, M., B\"oringer, H., 2009,
A\&A, 498, 361

\bibitem[2010]{pratt_10}
Pratt, G., et al., 2010, A\&A, 511, A85

\bibitem[1974]{press_74}
Press, W., Schechter, P., 1974, ApJ, 187, 425

\bibitem[1984]{primack 1984}
Primack, J., 1984, Enrico Fermi International Scoole of
Physics, Varenna, Italy.

\bibitem[2013]{rein-2013}
Reinhardt, C., et al., 2013, ApJ, 763, 127

\bibitem[2011]{ru-2011}
Rubakov, V., 2011, PhyU, 54, 633

\bibitem[2013]{ru-2013}
Ruel J., et al., 2013, arXiv:1311.4953

\bibitem[2014]{ruff-2014}
Ruffini  R., Arguelles, C., Rueda, J., 2014, in press.

\bibitem[2011]{salucci 2011}
Saliwanchik, B., et al. arXiv:1312.3015

\bibitem[2014]{sam 2014}
Samushia, L., et al., 2014, MNRAS, 439, 3504

\bibitem[2013]{saro 2013}
Saro, A., et al., 2013, arXiv:1312.2462

\bibitem[2013]{save 2013}
Savelainen, M., Valiviita, J., Walia, P., Rusak, S.,
Kurki--Suonio, H.,  2013, PhRvD., 88, 063010.

\bibitem[2013]{sawa 2013}
Sawala, T., Frenk, C., Crain, R., Jenkins, A., Schaye, J.,
Theuns, T., Zavala, J., 2013, MNRAS, 431, 1366

\bibitem[2013]{say 2013}
Sayers, J., et al., 2013, ApJ, 768, 177  

\bibitem[2013]{schnei 2013}
Schneider, A., Smith, R., Reed, D., 2013, MNRAS, 433, 1573

\bibitem[2013]{schn 2013}
Schneider, A., Anderhalden, D., Maccio, A., Diemand, J.,
2014, MNRAS, 441, L6

\bibitem[2014]{schu 2014}
Schultz, C., Onorbe, J., Abazajian, K., Bullock, J., 2014,
arXiv:1401.3769

\bibitem[2013]{souza 2005}
Souza, R., Mesinger, A., Ferrara, A., Haiman, Z., Perna, R.,
Yoshida, N., 2013, arXiv:1303.5060   >2keV

\bibitem[2011]{suhada}
Suhada, R., et al., 2012, A\&A, 537, 39, 1076,
arXiv:1111.0141

\bibitem[2002]{Tasit-02}
Tasitsiomi, A., Kravtsov, A., Gottl\"ober, S., Klypin, A.,
2004, ApJ, 607, 125

\bibitem[2013]{Tess.}
Teyssier, R., Pontzen, A., Dubois, Y., Read, J., 2013,
MNRAS, 429, 3068

\bibitem[2012]{Tollerud.}
Tollerud et al., 2012, ApJ, 752, 45

\bibitem[1979]{Trimain}
Trimain, S., Gunn, J., 1979, PRL, 42, 407

\bibitem[1984]{turner 1984}
Turner, M., Steigmann, G., Krauss, L., 1984, Phys.Rev.Let.,
52, 2090.

\bibitem[1993]{umemura}
Umemura, M., Loeb, A., Turner, E., 1993, ApJ,  419, 459

\bibitem[2013]{verde 2013}
Verde, L., Feeney, S., Mortlock, D., Peiris, H., 2013,
JCAP, 09, 13

\bibitem[2005]{viel 2005}
Viel, M., Lesgourgues, J., Haehnelt, M., Mattarese, S.,
Riotto, A., 2005, PhRvD, 71f3534

\bibitem[2013]{viel 2013}
Viel, M., Becker, G., Bolton, J., Haehnelt,M., 2013,
PhRvD, 88d3502

\bibitem[2006]{vikhlinin-06}
Vikhlinin, A., Kravtsov, A., Forman, W., Jones, C., Markevitch,
M., Murrey, S., Van Speybroeck, L.,, 2006, ApJ, 640, 691

\bibitem[2009]{vikhlinin}
Vikhlinin, A., et al., 2009, ApJ, 692, 1033

\bibitem[2013]{villa}
Villaescusa-Navarro, F., Marulli, F., Viel,M., Branchini, E.,
Castorina, E., Sefusatti, E., Saito, S. 2013, arXiv:1311.0866

\bibitem[2009]{walker 2009}
Walker, M., et al. 2009, ApJ., 704, 1274

\bibitem[2012]{walk 2012}
Walker, M., Penarrubia, J., 2011, ApJ,  742, 20   

\bibitem[2012]{walker 2012}
Walker, M. 2012, arXiv:1205.0311

\bibitem[2013]{wang 2013}
Wang, M.-Y., Croft, R., Peter, A., Zentner, A., Purcell, C.,
2013, 1309.7354

\bibitem[2008]{wise 2008}
Wise, J.,\,\&\,Abel. T.,2008, ApJ, 685, 40

\bibitem[2014]{wyman 14}
Wyman, M., Rudd, D., Vanderveld, R., Hu, W., 2014, PRL,
112, 051302

\bibitem[1970]{Zel'd-70}
Zel'dovich Ya.B.,  1970, A\&A, 5, 84

\bibitem[2003]{Zel'd}
Zel'dovich Ya.B., Novikov I.D., 1983, Structure and evolution
of the Universe, University of Chicago Press.

\bibitem[2003]{zhang 2006}
Zhang, Y., B\"oringer, H., Finoguenov, A., Ikebe, Y.,
Matsushita, K., Schueker, P., Guzzo, L., Collins, C., 2006,
A\&A, 456, 55

\end{thebibliography}
\end{document}